\documentclass[12pt,preprint]{aastex}
\usepackage{graphicx}
\usepackage{epstopdf}
\usepackage{amsmath}
\usepackage{threeparttable}
\begin{document}

\title{Can friction of the nova envelope account for the extra angular momentum loss in cataclysmic variables?}

\author{Wei-Min Liu$^{1,2,3}$ and Xiang-Dong Li$^{1,3}$}

\affil{$^{1}$Department of Astronomy, Nanjing University, Nanjing 210046, China}
\affil{$^{2}$Department of Physics, Shangqiu Normal University, Shangqiu 476000, China}
\affil{$^{3}$Key Laboratory of Modern Astronomy and Astrophysics (Nanjing University), Ministry of Education, Nanjing 210046, China}

\affil{$^{}$liuwmph@163.com; lixd@nju.edu.cn}

\begin{abstract}

It has been shown that the rate of angular momentum loss (AML) in cataclysmic variables (CVs) below the period gap is about 2.47 times that caused by gravitational radiation, suggesting extra AML mechanism besides gravitational radiation. Several potential mechanisms have been proposed but none of them has been verified. In this work we examine whether AML caused by friction between the expanding nova envelope and the donor star can account for the required AML rate. By adopting various expanding velocities of the envelope, we have calculated the evolution of CVs with typical initial parameters.  Our results show that this friction interaction unlikely solve the extra AML problem unless the expanding velocities are extremely low. Thus there should be a more efficient AML mechanism that plays a role in the CV evolution.
\end{abstract}


\section{Introduction}
Cataclysmic variables (CVs) are interacting binaries where a low-mass donor star is transferring material onto a white dwarf (WD) \citep[see][for reviews]{war95,rit10,kni11}. The orbital periods of most CVs are $\lesssim$1 day and the mass transfer is driven by orbital angular momentum loss (AML). In the standard model of CV evolution, AML is dominated by magnetic braking \citep[MB;][]{verb81,Rappaport1983} above the $\sim 2-3$ hour period gap,  while below the period gap, the donor star becomes fully convective so the effect of MB is ceased and only gravitational radiation \citep[GR;][]{landau75} works.

A study of the CV population led \citet{pat98} to suggest that the AML rate below the period gap may be higher than the traditional prescription, which predicts that most ($99\%$ of the total) CVs populate the short-period regime \citep{kolb93}.
\cite{kni11} reconstructed the full evolutionary path of CVs based on the observed mass-radius relationship of low-mass stars. Their best-fit results showed that the scaling factors $f_{\rm GR}$ and $f_{\rm MB}$ for the standard GR- and MB-induced AML rates respectively are $f_{\rm GR}=2.47(\pm0.02)$ below the period gap and $f_{\rm MB}=0.66(\pm0.05)$ above the period gap.
More recently, \cite{pala17} investigated the CV evolution using the effective temperatures of the WDs as a probe. They obtained a good fit with the observations provided that there is additional AML mechanism below the period gap. This mechanism could be the residual MB  when the donor has no radiative core \citep{pat98} or consequential AML due to mass loss along with the mass transfer process \citep[hereafter CAML;][]{sch16,nel16}. 

\cite{shao12} investigated three possible CAML mechanisms related to mass loss in CVs: (1) isotropic wind from the surface of the WDs \citep{king95}, (2) mass loss through the Lagrangian points $L_1$ or $L_2$ \citep{vanb98}, and (3) the formation of a circumbinary (CB) disk from the outflow \citep{vanden94,taam01}. They showed that neither isotropic wind nor outflow from the $L_1$ point can account for the extra 1.47 times GR-induced AML rate ($\dot{J}_{\rm GR}$), while outflow from the $L_2$ point or the formation of a CB disk may account for it, provided that $\sim (15-45)\%$ or $\sim (20-40)\%$ of the transferred material leaves the binary, respectively.  In reality, both isotropic wind and outflow may simultaneously play a role during the CV evolution, and \citet{liu16} found that the mass transfer in CVs with low-mass WDs becomes dynamically unstable in this case if the fraction mass loss in the form of a CB disk is $\sim (20-30)\%$.

Another important topic associated with the CV evolution is the absence of low-mass WDs in CVs \citep{de92,pol96,zor11,wij15}. To solve this problem  \cite{sch16} suggested an alternative empirical CAML model taking into account AML generated by the mass transfer in CVs. Assuming that the specific AM of the lost matter increases with decreasing WD mass, they carried out detailed CV population synthesis simulations. The results showed that this model could explain the discrepancy between the measurements and theoretical predictions of the average WD mass in CVs, as well as their orbital period distribution \citep{kolb99,kni06,gan09} and space density distribution \citep{ritter86,kolb93,pat98,Pre12} of CVs. However, the physical origin of this kind of CAML is not known, though it is likely to be related to mass loss during nova eruptions. \cite{sch98} investigated AML due to friction (hereafter FAML) between the expanding nova envelope and the donor star. They showed that the strength of FAML sensitively depends on the expanding velocity of the ejecta at the location of the donor, stronger for smaller expanding velocity. Interestingly, previous studies \citep{livio91,yar05} have shown that lower expanding velocities are expected in lower-mass WDs than in more massive ones. Therefore, FAML may present a potentially possible explanation for the empirical CAML law suggested by \cite{sch16}. However, it remains to examine whether FAML can explain the extra $1.47 \dot{J}_{\rm GR}$ below the period gap. This is the objective of our work.

The rest of this paper is organized as follows. In Section 2, we describe the input physics and the FAML model considered in the binary evolution calculations. The numerically calculated results are presented in Section 3 and compared with the analytic derivation in Section 4. We summarize our results in Section 5.

\section{Model}

We carried out binary evolutionary calculation of CVs using Modules for Experiments in Stellar Astrophysics (MESA) \citep{Paxton2011ApJS,Paxton2013ApJS,Paxton2015ApJS}. Generally during nova eruptions all the material accreted by the WD  is ejected. Therefore, there is not net mass accumulation of a WD in CVs. However, at the beginning of mass transfer, there could be a stage of stable hydrogen and helium burning depending on the mass ratio, so we include possible mass accumulation. The mass growth rate $\dot{M}_{\rm WD}$ of a WD can be described as follows,
\begin{equation}
\dot{M}_{\rm WD}=-\eta_{\rm H}\eta_{\rm He}{\dot M}_2,
\end{equation}
where $\eta_{\rm H}$ and $\eta_{\rm He}$ are the mass accumulation efficiencies for hydrogen and helium burning respectively, and $-{\dot M}_2$ is the mass transfer rate. We refer to \cite{hil16} and \cite{Ka2004} for the dependence of $\eta_{\rm H}$ and $\eta_{\rm He}$ on the WD mass and the mass transfer rate, respectively \cite[see also][for more details]{liu16}. The excess material is assumed to leave the binary at a rate of ($|{\dot M}_2|-{\dot M}_{\rm WD}$), taking away the specific AM of the WD. In particular, if the mass transfer rate is lower than $3\times 10^{-8}\,M_\sun$\,yr$^{-1}$, novae are supposed to take place where all the accreted material is ejected from the surface of the WD. In this case, besides AML due to mass loss, we also consider possible AML caused by frictional interaction (see below).

In our calculations, systematic AML such as GR \citep{landau75} and MB \citep{verb81} are included. We take Solar chemical abundance ($X = 0.70$, $Y = 0.28$, and $Z = 0.02$) for the donor star.
\subsection{Frictional angular momentum loss}
\cite{sch98} proposed that there may exist FAML between the expanding nova envelope and the secondary. They deduced the FAML expression based on the Bondi-Hoyle accretion model. Since the duration of nova eruptions is much shorter than the mass transfer time, we use their long-term continuous wind average to investigate its influence on the secular evolution of CVs. The mean specific AM $j_{\rm ej}$ of the ejected material during nova eruptions can be written as follows,
\begin{equation}
j_{\rm ej}=\left(q+\nu_{\rm FAML}\right)\frac{J}{M},
\end{equation}
with $q=M_2/M_{\rm WD}$ being the ratio of the donor mass and the WD mass,  and $J$ and $M$  the total AM and mass of the binary respectively. On the right-hand-side of Eq.~(2), the first term represents the specific AM of the WD carried by the expanding material, and the second the specific AM related to friction between the expanding envelope and the donor star. Here the parameter $\nu_{\rm FAML}$ represents the strength of FAML,
\begin{equation}
\nu_{\rm FAML}=\frac{(1+q)^2}{4q}\left(\frac{R_{\rm {L,2}}}{a}\right)^2\frac{v_{\rm rel}}{v_{\rm exp}},
\end{equation}
where $R_{\rm{L,2}}/a$ is the ratio of the Roche-lobe (RL) radius of the donor and the separation of the binary, $v_{\rm exp}$ the expansion velocity of the envelope from the WD, and $v_{\rm rel}$ the relative velocity of the expanding envelope to the donor star, $v_{\rm rel}=\sqrt{{v_{\rm sec}}^2+{v_{\rm exp}}^2}$, where $v_{\rm sec}=\sqrt{GM/a}$ is the orbital velocity about the WD. Since the donor star cannot spin the envelope up to velocities faster than corotation, we set an upper limit for FAML as \cite{sch98},
\begin{equation}
j_{\rm max}=j_1+\left(\frac{R_2}{a}\right)a^2\omega
\end{equation}
where $j_1$, $R_2$ and $\omega$ are the specific AM of the WD, the radius of the donor star and the angular velocity of the binary, respectively. Therefore, the actual specific AM of the ejecta including FAML is given by
\begin{equation}
j_{\rm ej}={\rm min}(j_{\rm ej}, j_{\rm max}).
\end{equation}

\section{Calculated results}

In our calculation, we take the initial WD mass to be $M_{\rm {WD,i}}=0.5$, 0.8 and 1.1\,$M_\odot$, and the corresponding initial donor star mass to be $M_2=(0.4,\ 0.5,\ 0.6)M_\sun$, $(0.6,\ 0.8,\ 1.0)M_\sun$, and $(0.6,\ 0.8,\ 1.0)M_\sun$, respectively. For the expanding velocity of the ejected envelope, we adopt $v_{\rm exp} = 40$, 80, and 200\,$\rm {km\,s}^{-1}$.

In Table 1 we list the calculated results without FAML (i.e., $\nu_{\rm FAML}=0$). These results can be used as reference for comparison with other models. We find that for $M_{\rm WD}= 0.5\,M_\sun$, when $q\gtrsim1$, the WD mass can grow to some extent. The reason is that these systems initially experience thermal-timescale mass transfer for some time. In other systems the mass transfer rates never exceed $3\times 10^{-8}\,M_\sun$\,yr$^{-1}$ during the whole evolution so they all experience novae. The minimum orbital period $P_{\rm {orb,min}}$ ($\lesssim 70$ min) depends on both the WD mass and the initial orbital period $P_{\rm {orb,i}}$: the larger $M_{\rm WD,i}$ and the shorter $P_{\rm {orb,i}}$, the longer $P_{\rm {orb,min}}$.

Tables 2-4 present the calculated results by taking FAML into account with $v_{\rm exp}=40$, 80, and 200 $\rm {km\,s}^{-1}$, respectively. FAML can slightly enhance the mass transfer rate compared with the no-FAML cases. For $M_{\rm {WD,i}}= 0.5\,M_\sun$ this leads to more efficient mass growth in the WDs, which increases with decreasing $v_{\rm exp}$. The minimum orbital period $P_{\rm {orb,min}}$ is not influenced by whether or not considering FAML.

In order to show more details in the evolution, Figs.~1-3 present three examples for different WD and donor masses. In Fig.~1, the initial parameters are $M_{\rm {2,i}} = 0.6\,M_\sun$, $M_{\rm {WD,i}}= 0.5\,M_\sun$ and $P_{\rm {orb,i}}=0.5$ day. The left panels display the evolution of the mass transfer rate (black solid line) and the orbital period (blue solid line). The right panels depict the evolution of the AML rate caused by GR (black solid line), the CAML rate (due to isotropic wind and FAML) (red solid line), and the ratio of  the CAML rate and the AML rate due to GR. The top panels correspond to the case without FAML. In the other three panels, from up to down, the values of $v_{\rm exp}$ are taken to be 40, 80 and 200\,$\rm {km\,s}^{-1}$, respectively.

In the left panels of Fig.~1 we show that, due to relatively large initial $q$, the system experiences a short thermal-timescale mass transfer phase during which the growth of the WD mass occurs, and when FAML starts to work, the mass transfer evolves discontinuously. The reason is that the amplified mass transfer causes the orbital period to increase, leading to temporary RL-detachment. This effect gradually declines when $v_{\rm exp}$ increases from 40\,$\rm {km\,s}^{-1}$ to 200\,$\rm {km\,s}^{-1}$.  In the right panels, variation in the mass transfer rate causes $\dot{J}_{\rm CAML}/\dot{J}_{\rm GR}$ to vary between very small ($\ll 1$) and large ($>10^4$) values. However, when the system has evolved across the period gap, the mass transfer rate decreases to low values ($\lesssim10^{-10}M_\sun\,{\rm yr^{-1}}$), and GR begins to dominate the evolution. We find that the CAML rate decreases with time and is considerably less than $\dot{J}_{\rm GR}$ during this stage.
The FAML effect becomes weaker with increasing $v_{\rm exp}$. When $v_{\rm exp}=$ 200 $\rm {km\,s}^{-1}$, the evolution below the period gap is very close to that without FAML.

In Figs.~2 and 3, the initial WD masses are taken to be $M_{\rm {WD,i}}= 0.8\,M_\sun$ and $1.1\,M_\sun$, respectively. Due to the relatively small $q$ value, the mass transfer rate has no chance to exceed $\sim 3\times 10^{-8}\,M_\sun$\,yr$^{-1}$ and nova eruptions prevent the mass growth of the WDs. In the case of FAML, the mass transfer rate is enhanced by several times for $v_{\rm exp}=$ 40 $\rm {km\,s}^{-1}$, compared with the case of no FAML, but still cannot enter the steady hydrogen burning regime.
The values of $\dot{J}_{\rm CAML}/\dot{J}_{\rm GR}$ are similar for a given $v_{\rm exp}$, and all significantly less than unity.

\section{Discussion}
In last section, we examine whether FAML can solve the extra $1.47\dot{J}_{\rm GR}$ AML  problem for CVs below the period gap by calculating the secular evolution of CVs.
In this section, we first attempt to derive possible constraint on the extra AML mechanism in an analytical way. For the standard evolution of CVs, we assume that all the accreted mass is lost from the binary during nova eruptions, taking the specific AM of the WD. We can obtain the mass transfer rate as follows \citep{Rappaport1983}
\begin{equation}
-\frac{\dot{M}_{2}}{M_{2}}=\frac{\frac{1}{2}\left(\frac{\dot{R_{2}}}{R_{2}}\right)_{\rm ev,th}
-\left(\frac{\dot{J}_{\rm sys}}{J}\right)}{\frac{5}{6}+\frac{\zeta}{2}-\frac{q}{3(1+q)}-\frac{q^2}{1+q}},
\end{equation}
where $(\dot{R_{2}}/R_{2})_{\rm ev,th}$ denotes the change in the donor star's radius due to thermal or nuclear evolution, $\dot{J}_{\rm sys}$ is the systematic AML rate, and $\zeta$ is the adiabatic mass-radius exponent of the donor, namely $R_2\propto M_2^{\zeta}$. For low-mass main-sequence (MS) stars, the change in the stellar radius due to evolution can be neglected, i.e.,$(\frac{\dot{R}_{2}}{R_2})_{\rm ev,th}\simeq 0$. For CVs below the period gap, let $\dot{J}_{\rm sys}=2.47\dot{J}_{\rm GR}$, Eq.~(6) can be rewritten to be
\begin{equation}
\frac{\dot{M}_{2}}{M_{2}}=\frac{\left(\frac{2.47\dot{J}_{\rm GR}}{J}\right)}{\frac{5}{6}+\frac{\zeta}{2}-\frac{q}{3(1+q)}-\frac{q^2}{1+q}}.
\end{equation}
From Eq.~(2) we have
\begin{equation}
\frac{\dot{J}_{\rm CAML}}{J}=\left(q+\nu_{\rm FAML}\right)\frac{M_2}{M}\frac{\dot{M}_2}{M_2},
\end{equation}
Let $\dot{J}_{\rm CAML}=1.47\dot{J}_{\rm GR}$, we obtaiin
\begin{equation}
\frac{1.47\dot{J}_{\rm GR}}{J}=\left(q+\nu_{\rm FAML}\right)\frac{M_2}{M}\frac{\dot{M}_2}{M_2},
\end{equation}
Combining Eqs.~(2), (3), (7), and (9), and using the empirical formula for the RL radius \citep{sch98}
\begin{equation}
\frac{R_{\rm {L,2}}}{a}=\left[\frac{8q}{81(1+q)}\right]^{1/3}
\end{equation}
we finally get
\begin{equation}
x=\frac{(0.496+0.297\zeta)-\frac{q(1.47+3q)}{2.47\times3(1+q)}}{\frac{(1+q)}{4}\left[\frac{8q}{81(1+q)}\right]^{2/3}}
\end{equation}
where $x=v_{\rm rel}/v_{\rm exp}$. Eq.~(11) presents the required relation between $x$ and $q$ if FAML can account for the extra $1.47\dot{J}_{\rm GR}$ AML below the period gap. It is shown in Fig.~4 with the black solid line based on fit of the calculations for $M_{\rm 2, i}=0.6\,M_\odot$ and $P_{\rm {orb,i}}$ = 0.5 day and $\zeta\simeq 0.6$ \citep{liu16}.  In comparison, we also plot the calculated $x$ as a function of $q$ during the evolution of CVs. The red, blue and purple lines correspond to $M_{\rm WD,i} = 0.5$, 0.8 and $1.1\,M_\odot$, and  the solid and dotted lines correspond to $v_{\rm exp}= 40$ and $200\,\rm {km\,s}^{-1}$, respectively.
Obviously the calculated $x$ is much smaller than required, implying that the adopted values of $v_{\rm exp}$ are still too high.

In general, the expanding velocity of the ejected envelope for classical novae may exceed $1,000\,{\rm {km\,s}^{-1}}$ \citep{bode08}. \cite{yar05} showed the mean expanding velocity of the ejected envelope are related to the mass and core temperature of the WDs and the mass transfer rate, ranging from less than $100\, {\rm {km\,s}^{-1}}$ to over $3,000\, {\rm {km\,s}^{-1}}$ if the mass transfer rate  $\gtrsim10^{-12}M_\sun\,{\rm yr^{-1}}$. Their simulations also indicate that the expanding velocities are inversely correlated with the WD masses.
Thus the adopted values of $v_{\rm exp}$ in our calculations actually underestimate $v_{\rm exp}$. So we conclude that the FAML mechanism seems unable to explain the extra AML below the period gap for CVs.

We then discuss the caveats and uncertainties in our work. \cite{sch16} assumed that triggering dynamically unstable mass transfer in CVs with low-mass WDs may be the result of a slow nova resembling a common envelope phase. While the nova eruptions are discontinuous events, our adoption of the long-term continuous wind interaction may not exactly reflect their real influence on the binary evolution. In addition, the drag force for the Bondi-Hoyle accretion in the common envelope is subject to substantial uncertainties. There is a dimensional parameter $c_{\rm drag}$ in its expression \citep[Eq.~(31) in][]{sch98} with its default value set to be 2. However, \cite{kley95} pointed out that radiation pressure may reduce $c_{\rm drag}$ by a factor of 20. In Table 5, we present the calculated results by setting this parameter to be 0.4 and 0.1 with $v_{\rm exp}=40\,{\rm {km\,s}^{-1}}$, denoted by cases A and B, respectively. For both cases, we find that all results are quite similar with those in Tables   1 and 3, suggesting that the overall effect of FAML compared with GR is considerably small.

It should also be mentioned that the predicted minimum orbital periods (around $65-70$ mins) of CVs are substantially shorter than the observed ones (around $75-80$ mins).  These results are possibly related to the inadequacy of the AML models which affect the evolutionary time of the donor structures, thus leading the mass-radius relation to reverse earlier or later.
Setting the MB- and GR-induced AML rates to be 0.66 and 2.47 times the standard values above and below the period gap respectively, we re-calculate the CV evolution and present the results in Table 5, denoted to be case C. We find that the calculated minimum orbital periods are in accordance with the observed distribution. This further suggests that effect of FAML on the orbital period evolution is very limited.

\section{Conclusions}
We summarize our results as follows.

(1) FAML in the form of continuous wind interaction seems unable to account for the extra AML apart from GR below the period gap for CVs, except that the nova envelope has extremely low expanding velocities.

(2) FAML has very limited influence on the minimum orbital period distribution of CVs.

Our results imply that there should be a more efficient mechanism that plays a role in the CV evolution. The potential candidates could be residual MB and a CB disk. Thus infrared observations of CVs will be crucial in testing the latter idea.

\acknowledgments We are grateful to the referee for his/her valuable comments that helped improve this paper. This work was supported by the National Key Research and Development Program of China (2016YFA0400803), the Natural Science Foundation of China under grant Nos. 11333004, 11773015, and 11573016, Project U1838201 supported by NSFC and CAS, and the Program for Innovative Research Team (in Science and Technology) at the University of Henan Province.

\begin{table}
\begin{center}
\caption{The calculated results for the traditional evolution of CVs.}
\centering
\begin{tabular}{ccccccc}
\hline\hline
$M_2$ & $M_{\rm {WD,i}}$ &$M_{\rm {WD,f}}$ &$P_{\rm {orb,i}}$ &$P_{\rm {orb,min}}$ \\
($M_\odot$) &($M_\odot$) &($M_\odot$) &(days)&(mins) \\
\hline
0.4   &0.5   &0.5   &0.316   &67.55   \\
0.4   &0.5   &0.5   &0.501   &67.52   \\
0.4   &0.5   &0.5   &0.794   &67.42   \\
\hline
0.5   &0.5   &0.545   &0.316   &67.85   \\
0.5   &0.5   &0.548   &0.501   &67.84   \\
0.5   &0.5   &0.547   &0.794   &67.76   \\
\hline
0.6   &0.5   &0.534   &0.316   &67.79   \\
0.6   &0.5   &0.537   &0.501   &67.79   \\
0.6   &0.5   &0.538   &0.794   &67.74   \\
\hline
0.6   &0.8   &0.8   &0.316   &69.18   \\
0.6   &0.8   &0.8   &0.501   &69.16   \\
0.6   &0.8   &0.8   &1.0    &68.91   \\
\hline
0.8   &0.8   &0.8   &0.316   &69.18   \\
0.8   &0.8   &0.8   &0.501   &69.16   \\
0.8   &0.8   &0.8   &1.0    &68.97   \\
\hline
1.0   &0.8   &0.8   &0.316   &69.19  \\
1.0   &0.8   &0.8   &0.501   &69.18   \\
1.0   &0.8   &0.8   &1.0    &68.96  \\
\hline
0.6   &1.1   &1.1   &0.316   &70.35   \\
0.6   &1.1   &1.1   &0.501   &70.32  \\
0.6   &1.1   &1.1   &1.0    &70.05  \\
\hline
0.8   &1.1   &1.1   &0.316   &70.35  \\
0.8   &1.1   &1.1   &0.501   &70.32   \\
0.8   &1.1   &1.1   &1.0    &70.07  \\
\hline
1.0   &1.1   &1.1   &0.316   &70.35   \\
1.0   &1.1   &1.1   &0.501   &70.34   \\
1.0   &1.1   &1.1   &1.0    &70.03  \\
\hline\hline
\end{tabular}
\end{center}
\end{table}
\clearpage

\begin{table}
\begin{center}
\caption{The calculated results for the evolution of CVs with FAML for $v_{\rm exp}$ = 40 ${\rm km}\,{\rm {s}^{-1}}$.}
\centering
\begin{tabular}{ccccccc}
\hline\hline
$M_2$ & $M_{\rm {WD,i}}$ &$M_{\rm {WD,f}}$ &$P_{\rm {orb,i}}$ &$P_{\rm {orb,min}}$ &$v_{\rm exp}$\\
($M_\odot$) &($M_\odot$) &($M_\odot$) &(days)&(mins)&${\rm km}\,{\rm {s}^{-1}}$ \\
\hline
0.4   &0.5   &0.541   &0.316   &69.12   &40   \\
0.4   &0.5   &0.539   &0.501   &69.08   &40   \\
0.4   &0.5   &0.538   &0.794   &68.97   &40   \\
\hline
0.5   &0.5   &0.549   &0.316   &69.18   &40   \\
0.5   &0.5   &0.553   &0.501   &69.17   &40   \\
0.5   &0.5   &0.549   &0.794   &69.07   &40   \\
\hline
0.6   &0.5   &0.584   &0.316   &69.37   &40   \\
0.6   &0.5   &0.586   &0.501   &69.36   &40   \\
0.6   &0.5   &0.585   &0.794   &69.30   &40   \\
\hline
0.6   &0.8   &0.8   &0.316   &70.36   &40   \\
0.6   &0.8   &0.8   &0.501   &70.33   &40   \\
0.6   &0.8   &0.8   &1.0     &70.12   &40   \\
\hline
0.8   &0.8   &0.8   &0.316   &70.36   &40   \\
0.8   &0.8   &0.8   &0.501   &70.34   &40   \\
0.8   &0.8   &0.8   &1.0    &70.12   &40   \\
\hline
1.0   &0.8   &0.8   &0.316   &70.37   &40   \\
1.0   &0.8   &0.8   &0.501   &70.36   &40   \\
1.0   &0.8   &0.8   &1.0    &70.14  &40  \\
\hline
0.6   &1.1   &1.1   &0.316   &71.44   &40   \\
0.6   &1.1   &1.1   &0.501   &71.42   &40   \\
0.6   &1.1   &1.1   &1.0     &71.16   &40   \\
\hline
0.8   &1.1   &1.1   &0.316   &71.45   &40   \\
0.8   &1.1   &1.1   &0.501   &71.42   &40   \\
0.8   &1.1   &1.1   &1.0    &71.12    &40   \\
\hline
1.0   &1.1   &1.1   &0.316   &71.46   &40   \\
1.0   &1.1   &1.1   &0.501   &71.44   &40   \\
1.0   &1.1   &1.1   &1.0    &71.14    &40   \\
\hline\hline
\end{tabular}
\end{center}
\end{table}
\clearpage

\begin{table}
\begin{center}
\caption{The calculated results for the evolution of CVs for $v_{\rm exp}$ = 80 ${\rm km}\,{\rm {s}^{-1}}$.}
\centering
\begin{tabular}{ccccccc}
\hline\hline
$M_2$ & $M_{\rm {WD,i}}$ &$M_{\rm {WD,f}}$ &$P_{\rm {orb,i}}$ &$P_{\rm {orb,min}}$ &$v_{\rm exp}$\\
($M_\odot$) &($M_\odot$) &($M_\odot$) &(days)&(mins)&${\rm km}\,{\rm {s}^{-1}}$ \\
\hline
0.4   &0.5   &0.5024   &0.316   &68.20   &80   \\
0.4   &0.5   &0.5023   &0.501   &68.17   &80   \\
0.4   &0.5   &0.5021   &0.794   &68.07   &80   \\
\hline
0.5   &0.5   &0.5461   &0.316   &68.48   &80   \\
0.5   &0.5   &0.5464   &0.501   &68.46   &80   \\
0.5   &0.5   &0.5472   &0.794   &68.38   &80   \\
\hline
0.6   &0.5   &0.5795   &0.316   &68.67   &80   \\
0.6   &0.5   &0.5802   &0.501   &68.66   &80   \\
0.6   &0.5   &0.5803   &0.794   &68.60   &80   \\
\hline
0.6   &0.8   &0.8   &0.316   &69.74   &80   \\
0.6   &0.8   &0.8   &0.501   &69.72   &80   \\
0.6   &0.8   &0.8   &1.0    &69.52   &80   \\
\hline
0.8   &0.8   &0.8   &0.316   &69.75   &80   \\
0.8   &0.8   &0.8   &0.501   &69.73   &80   \\
0.8   &0.8   &0.8   &1.0    &69.52   &80   \\
\hline
1.0   &0.8   &0.8   &0.316   &69.76   &80   \\
1.0   &0.8   &0.8   &0.501   &69.75   &80   \\
1.0   &0.8   &0.8   &1.0    &69.54  &80   \\
\hline
0.6   &1.1   &1.1   &0.316   &70.88   &80   \\
0.6   &1.1   &1.1   &0.501   &70.85   &80   \\
0.6   &1.1   &1.1   &1.0     &70.61   &80   \\
\hline
0.8   &1.1   &1.1   &0.316   &70.88   &80   \\
0.8   &1.1   &1.1   &0.501   &70.85   &80   \\
0.8   &1.1   &1.1   &1.0    &70.54    &80   \\
\hline
1.0   &1.1   &1.1   &0.316   &70.88   &80   \\
1.0   &1.1   &1.1   &0.501   &70.87   &80   \\
1.0   &1.1   &1.1   &1.0    &70.59    &80   \\
\hline\hline
\end{tabular}
\end{center}
\end{table}
\clearpage

\begin{table}
\begin{center}
\caption{The calculated results for the evolution of CVs for $v_{\rm exp}$ = 200 ${\rm km}\,{\rm {s}^{-1}}$.}
\centering
\begin{tabular}{ccccccc}
\hline\hline
$M_2$ & $M_{\rm {WD,i}}$ &$M_{\rm {WD,f}}$ &$P_{\rm {orb,i}}$ &$P_{\rm {orb,min}}$ &$v_{\rm exp}$ \\
($M_\odot$) &($M_\odot$) &($M_\odot$) &(days)&(mins)&${\rm km}\,{\rm {s}^{-1}}$ \\
\hline
0.4   &0.5   &0.5006   &0.316   &67.82   &200   \\
0.4   &0.5   &0.5006   &0.501   &67.79   &200   \\
0.4   &0.5   &0.5005   &0.794   &67.68   &200   \\
\hline
0.5   &0.5   &0.5465   &0.316   &68.11   &200   \\
0.5   &0.5   &0.5471   &0.501   &68.09   &200   \\
0.5   &0.5   &0.5478   &0.794   &68.02   &200   \\
\hline
0.6   &0.5   &0.5376   &0.316   &68.06   &200   \\
0.6   &0.5   &0.5386   &0.501   &68.05   &200   \\
0.6   &0.5   &0.5382   &0.794   &68.00   &200   \\
\hline
0.6   &0.8   &0.8   &0.316   &69.41   &200   \\
0.6   &0.8   &0.8   &0.501   &69.39   &200   \\
0.6   &0.8   &0.8   &1.0    &69.11   &200   \\
\hline
0.8   &0.8   &0.8   &0.316   &69.41   &200   \\
0.8   &0.8   &0.8   &0.501   &69.39   &200   \\
0.8   &0.8   &0.8   &1.0    &69.17   &200   \\
\hline
1.0   &0.8   &0.8   &0.316   &69.42   &200   \\
1.0   &0.8   &0.8   &0.501   &69.41   &200   \\
1.0   &0.8   &0.8   &1.0    &69.20  &200   \\
\hline
0.6   &1.1   &1.1   &0.316   &70.56   &200   \\
0.6   &1.1   &1.1   &0.501   &70.54   &200   \\
0.6   &1.1   &1.1   &1.0     &70.26   &200   \\
\hline
0.8   &1.1   &1.1   &0.316   &70.56   &200   \\
0.8   &1.1   &1.1   &0.501   &70.54   &200   \\
0.8   &1.1   &1.1   &1.0    &70.28    &200   \\
\hline
1.0   &1.1   &1.1   &0.316   &70.57   &200   \\
1.0   &1.1   &1.1   &0.501   &70.55   &200   \\
1.0   &1.1   &1.1   &1.0    &70.28    &200   \\
\hline\hline
\end{tabular}
\end{center}
\end{table}
\clearpage

\begin{table}
\begin{center}
\caption{Examples of CV evolution in various cases.}
\centering
\begin{tabular}{ccccccc}
\hline\hline
$M_2$ & $M_{\rm {WD,i}}$ &$M_{\rm {WD,f}}$ &$P_{\rm {orb,i}}$ &$P_{\rm {orb,min}}$ &$v_{\rm exp}$ &Case\\
($M_\odot$) &($M_\odot$) &($M_\odot$) &(days)&(mins) &${\rm km}\,{\rm {s}^{-1}}$ \\
\hline
0.6   &0.5   &0.5376   &0.316   &68.05  &40 &A \\
0.6   &0.5   &0.5387   &0.501   &68.04  &40 &A \\
0.6   &0.5   &0.5385   &0.794   &67.99  &40 &A \\
\hline
0.6   &0.8   &0.8   &0.316   &69.40  &40 &A \\
0.6   &0.8   &0.8   &0.501   &69.37  &40 &A \\
0.6   &0.8   &0.8   &1.0    &69.17   &40 &A \\
\hline
0.6   &1.1   &1.1   &0.316   &70.55  &40 &A \\
0.6   &1.1   &1.1   &0.501   &70.53  &40 &A \\
0.6   &1.1   &1.1   &1.0    &70.13   &40 &A \\
\hline
0.6   &0.5   &0.5377   &0.316   &67.87  &40 &B \\
0.6   &0.5   &0.5354   &0.501   &67.84  &40 &B \\
0.6   &0.5   &0.5358   &0.794   &67.78  &40 &B \\
\hline
0.6   &0.8   &0.8   &0.316   &69.23  &40 &B \\
0.6   &0.8   &0.8   &0.501   &69.21  &40 &B \\
0.6   &0.8   &0.8   &1.0    &69.01   &40 &B \\
\hline
0.6   &1.1   &1.1   &0.316   &70.40  &40 &B \\
0.6   &1.1   &1.1   &0.501   &70.38  &40 &B \\
0.6   &1.1   &1.1   &1.0    &70.11   &40 &B \\
\hline
0.6   &0.5   &0.5386   &0.501   &74.63  &N/A &C \\
0.6   &0.8   &0.8      &0.501   &76.56  &N/A &C \\
0.6   &1.1   &1.1      &0.501   &78.22 &N/A &C \\
\hline\hline
\end{tabular}
\end{center}
\end{table}
\clearpage

\begin{figure}
\centering
\includegraphics[scale=0.25]{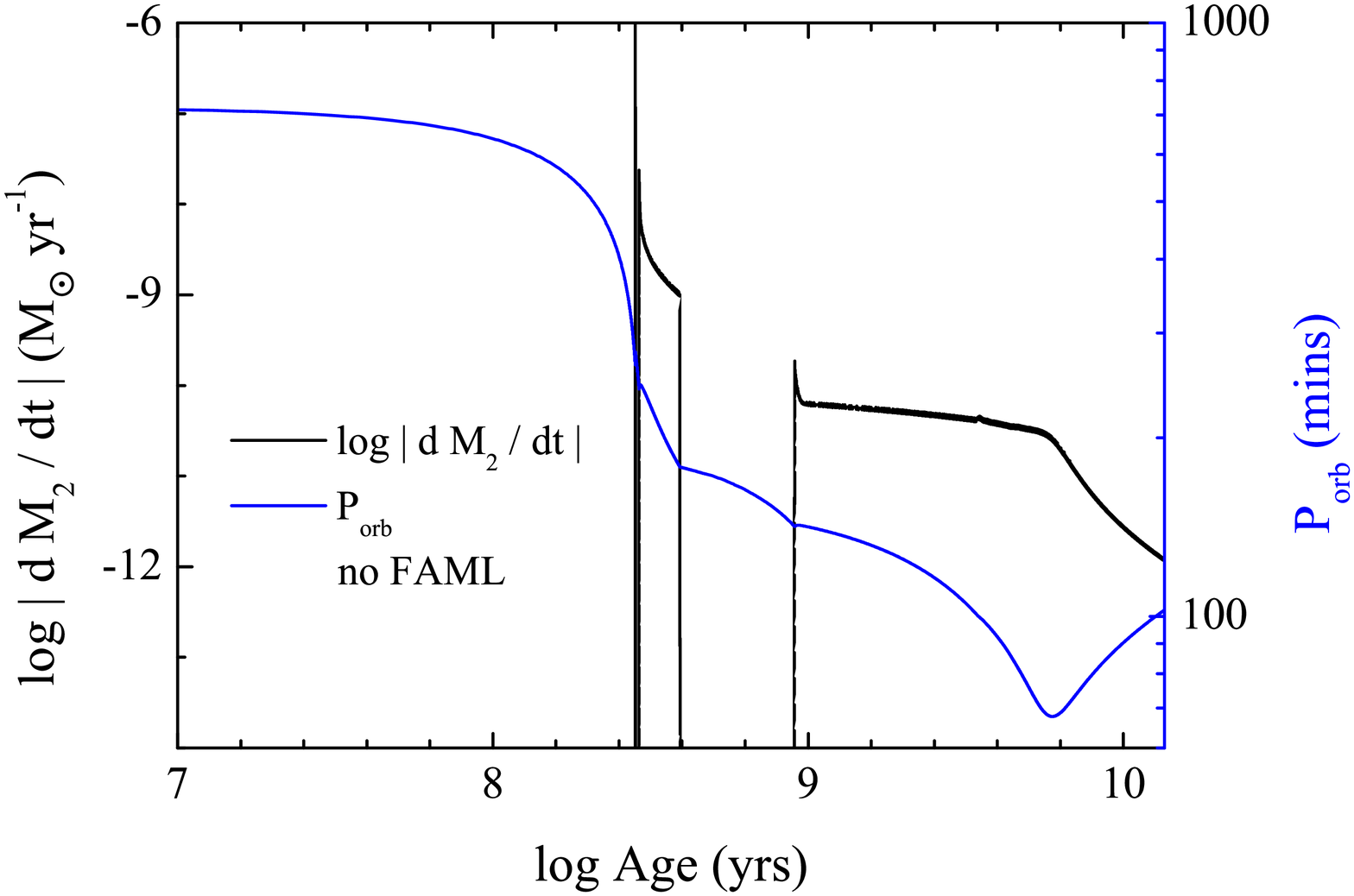}\includegraphics[scale=0.25]{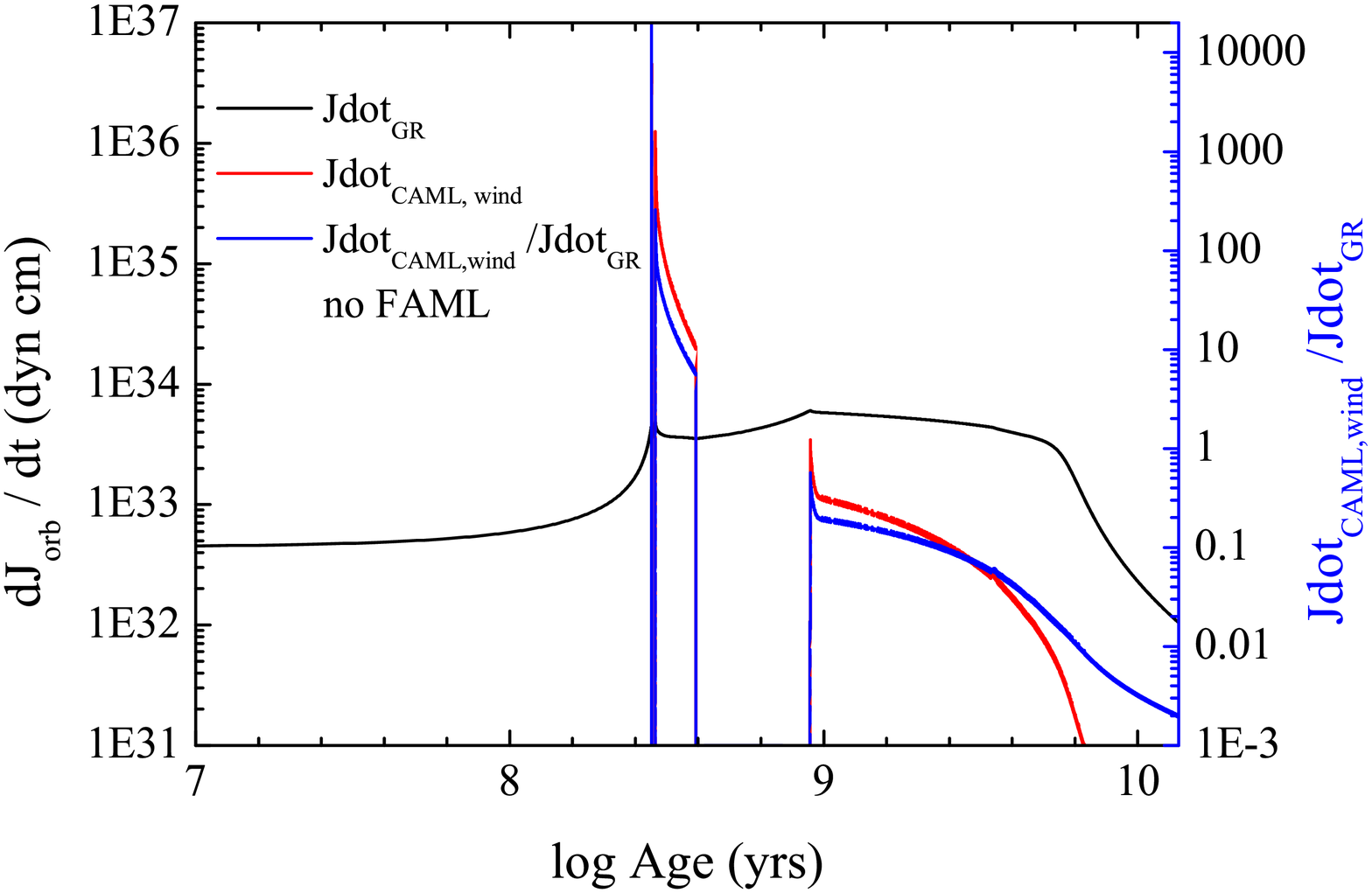}
\includegraphics[scale=0.25]{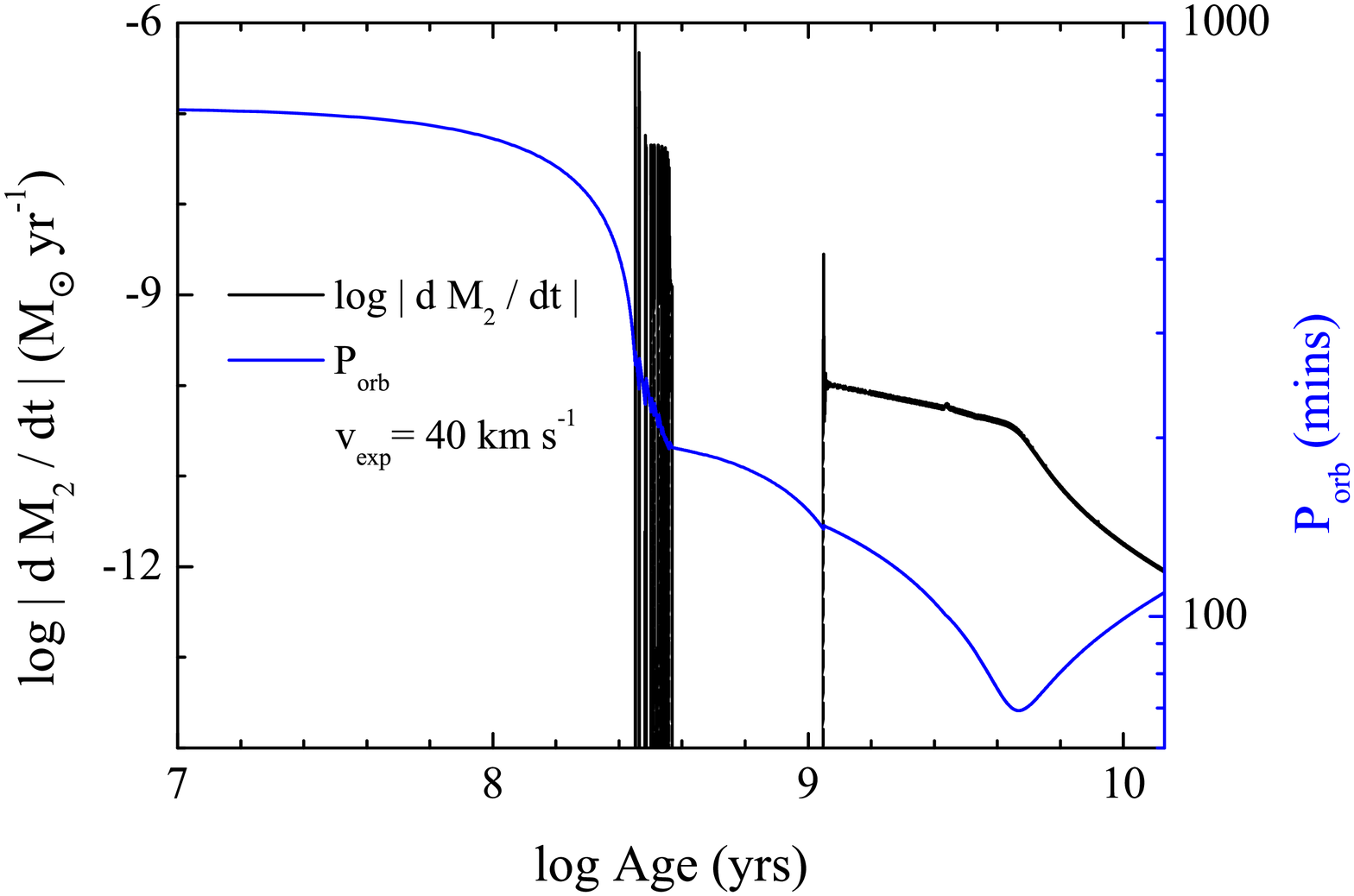}\includegraphics[scale=0.25]{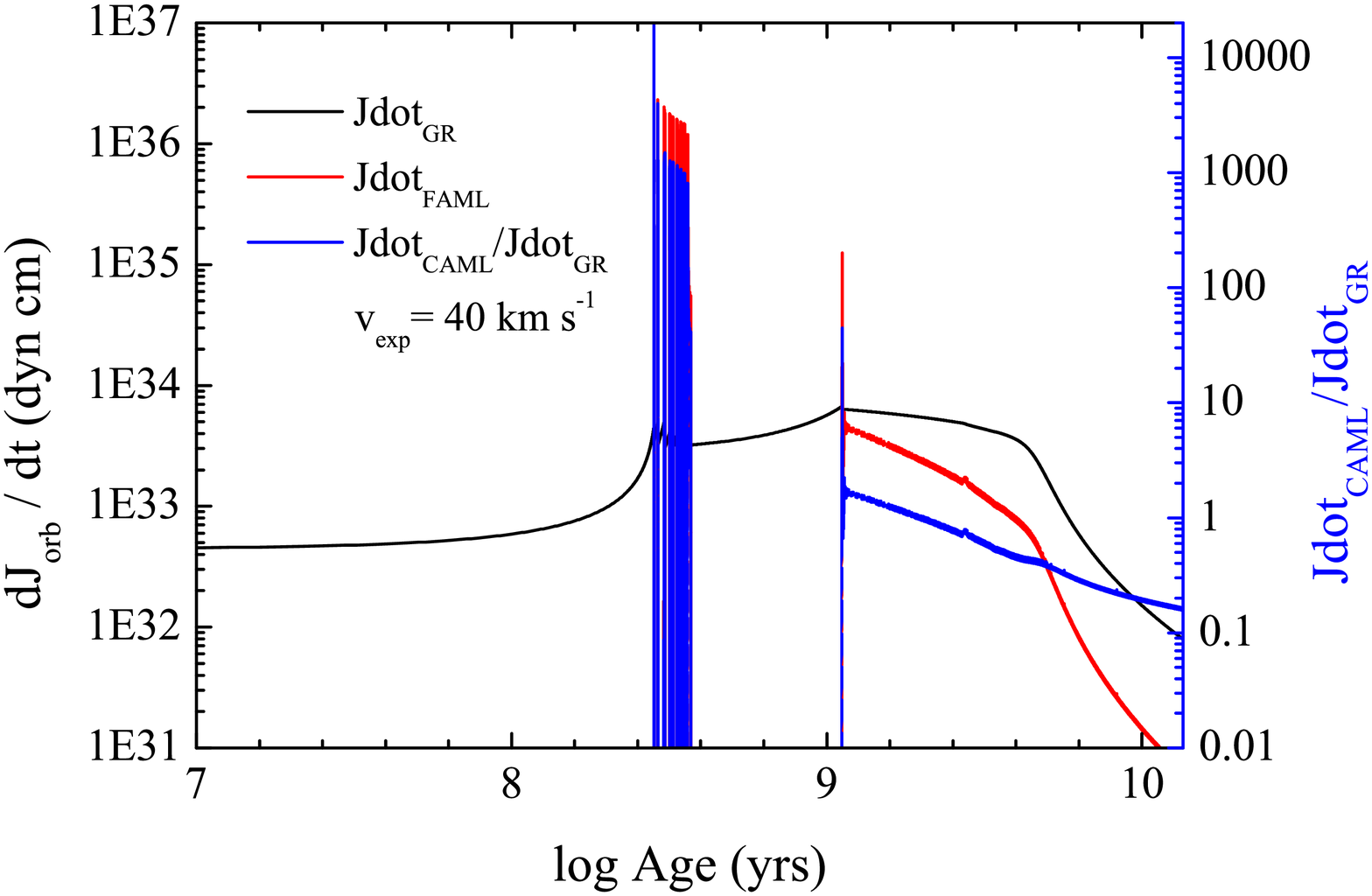}
\includegraphics[scale=0.25]{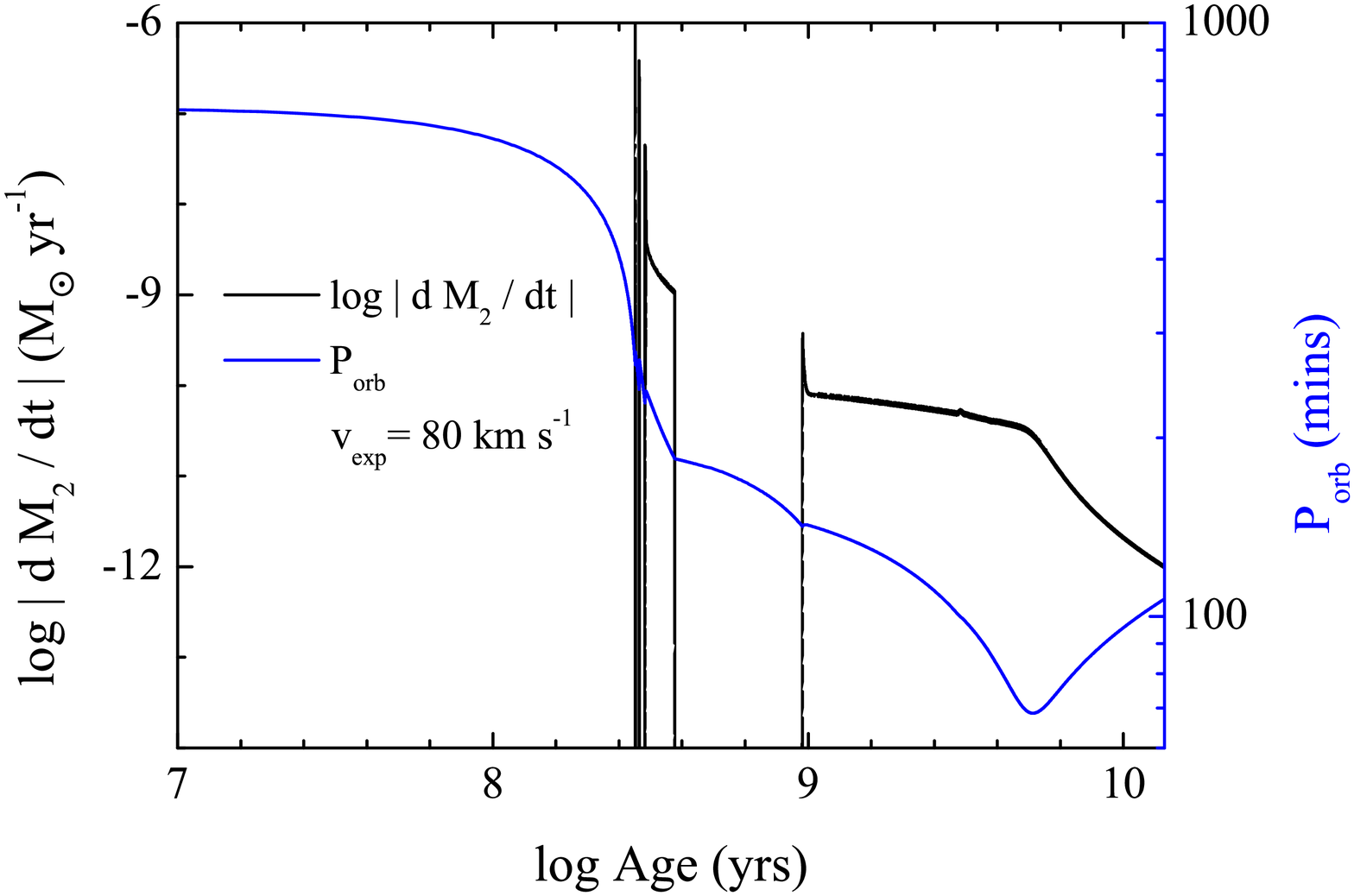}\includegraphics[scale=0.25]{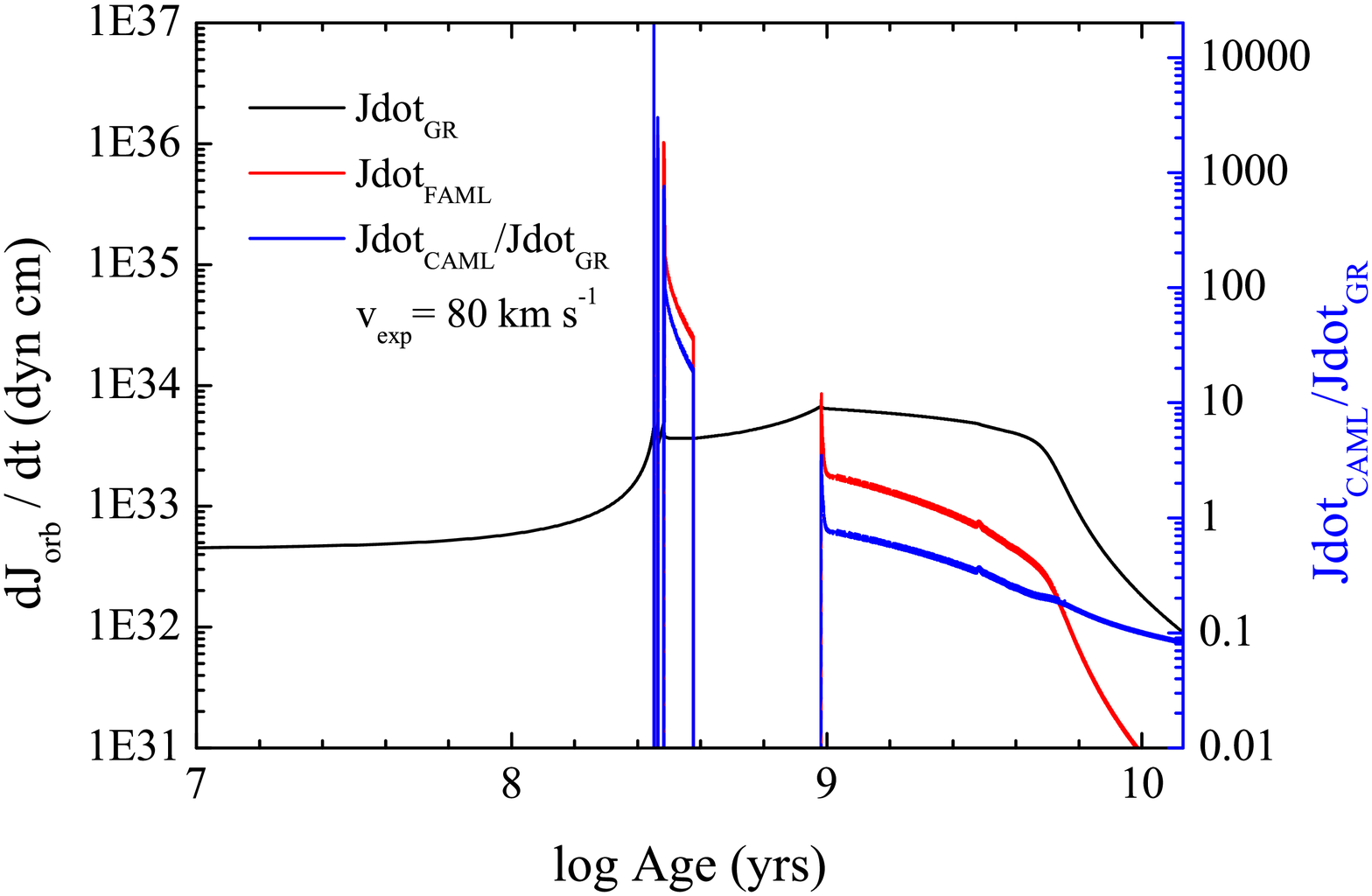}
\includegraphics[scale=0.25]{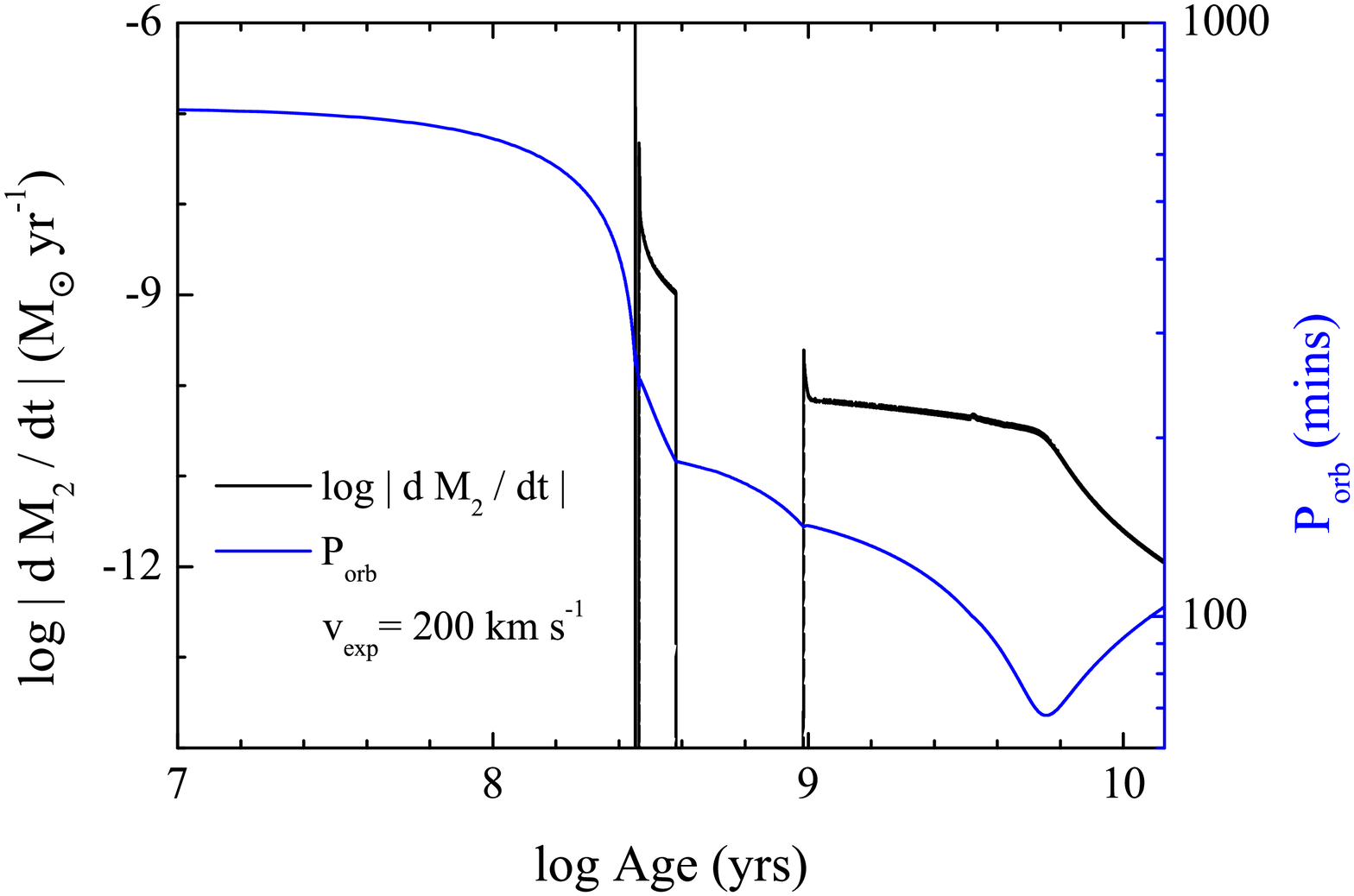}\includegraphics[scale=0.25]{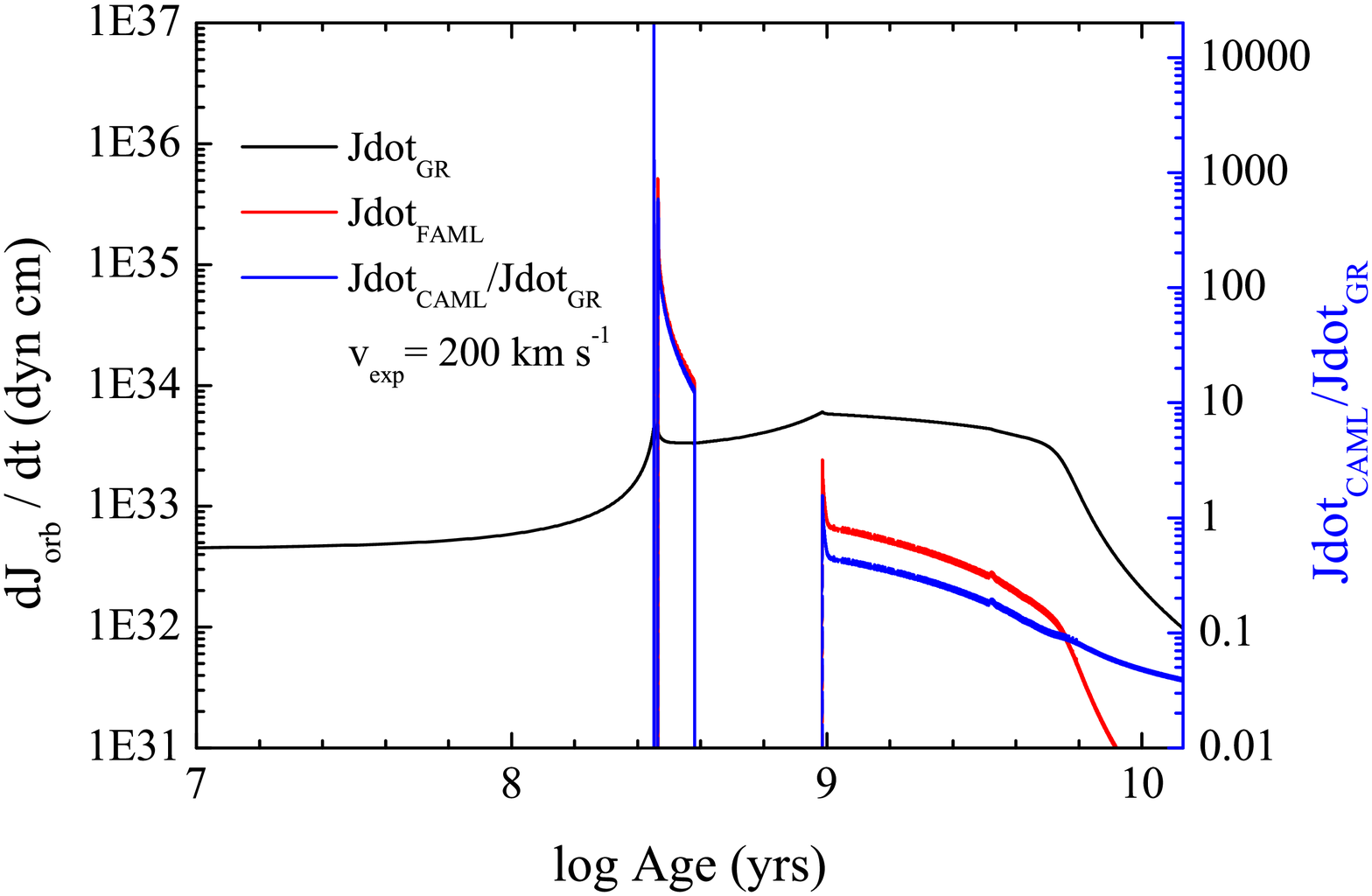}
\caption{The evolutionary tracks of a CV with $M_{\rm {2,i}}$ = 0.6 $M_\odot$, $M_{\rm {WD,i}}$ = 0.5 $M_\odot$ and $P_{\rm orb,i}$ = 0.5 day. The top panels show the traditional CV evolution. In other panels FAML is included with $v_{\rm exp}$ = 40, 80, and 200 ${\rm km}\,{\rm {s}^{-1}}$ from up to down. In the left panels, the black and blue lines represent the evolution of the mass transfer rate and the orbital period. In the right panels, the black, red and blue lines denote the AML rate due to GR, the CAML rate, and the ratio of the AML rates caused by CAML and GR, respectively.}
\label{fig:subfig}

\end{figure}

\clearpage

\begin{figure}
\centering
\includegraphics[scale=0.25]{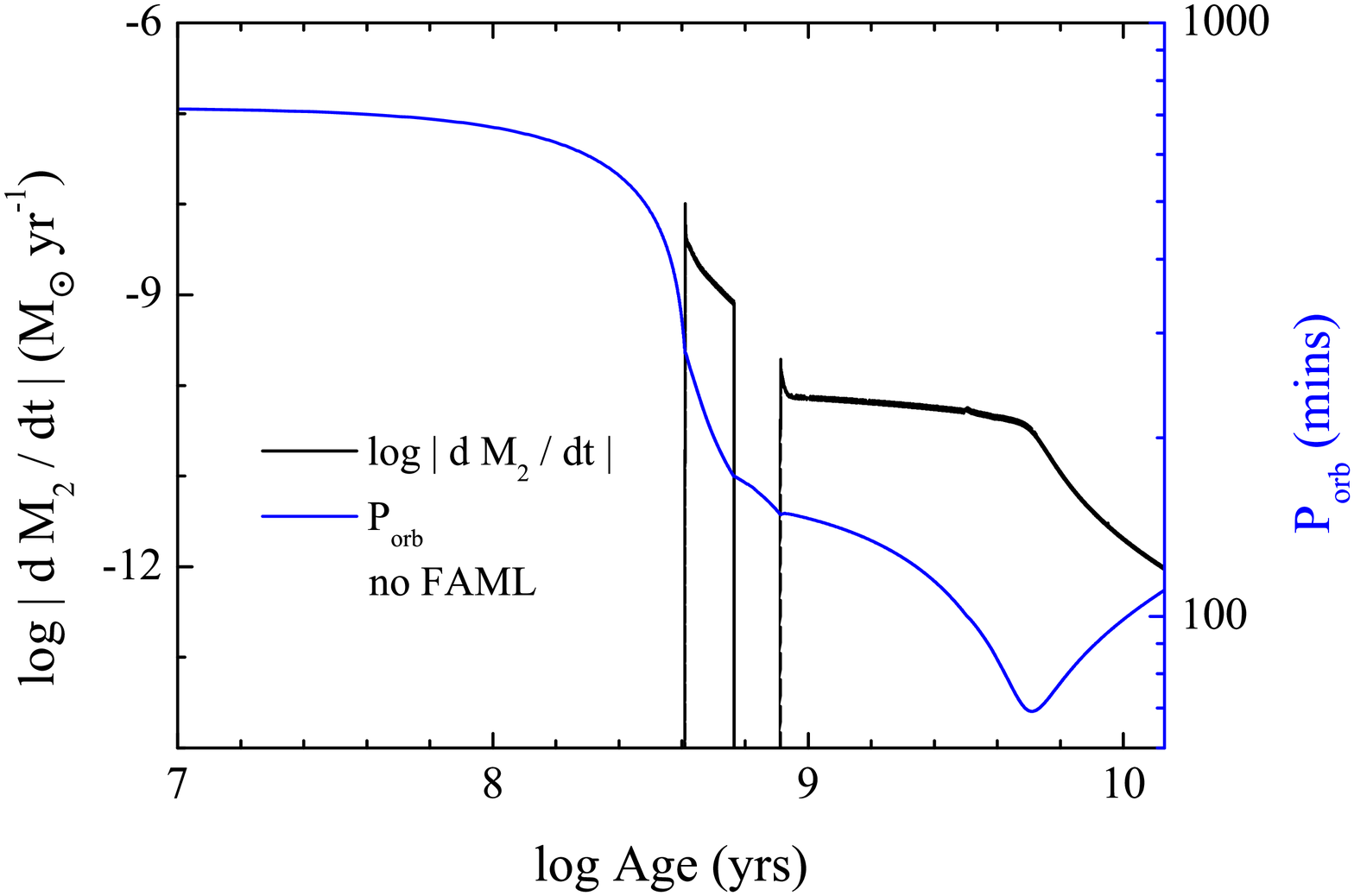}\includegraphics[scale=0.25]{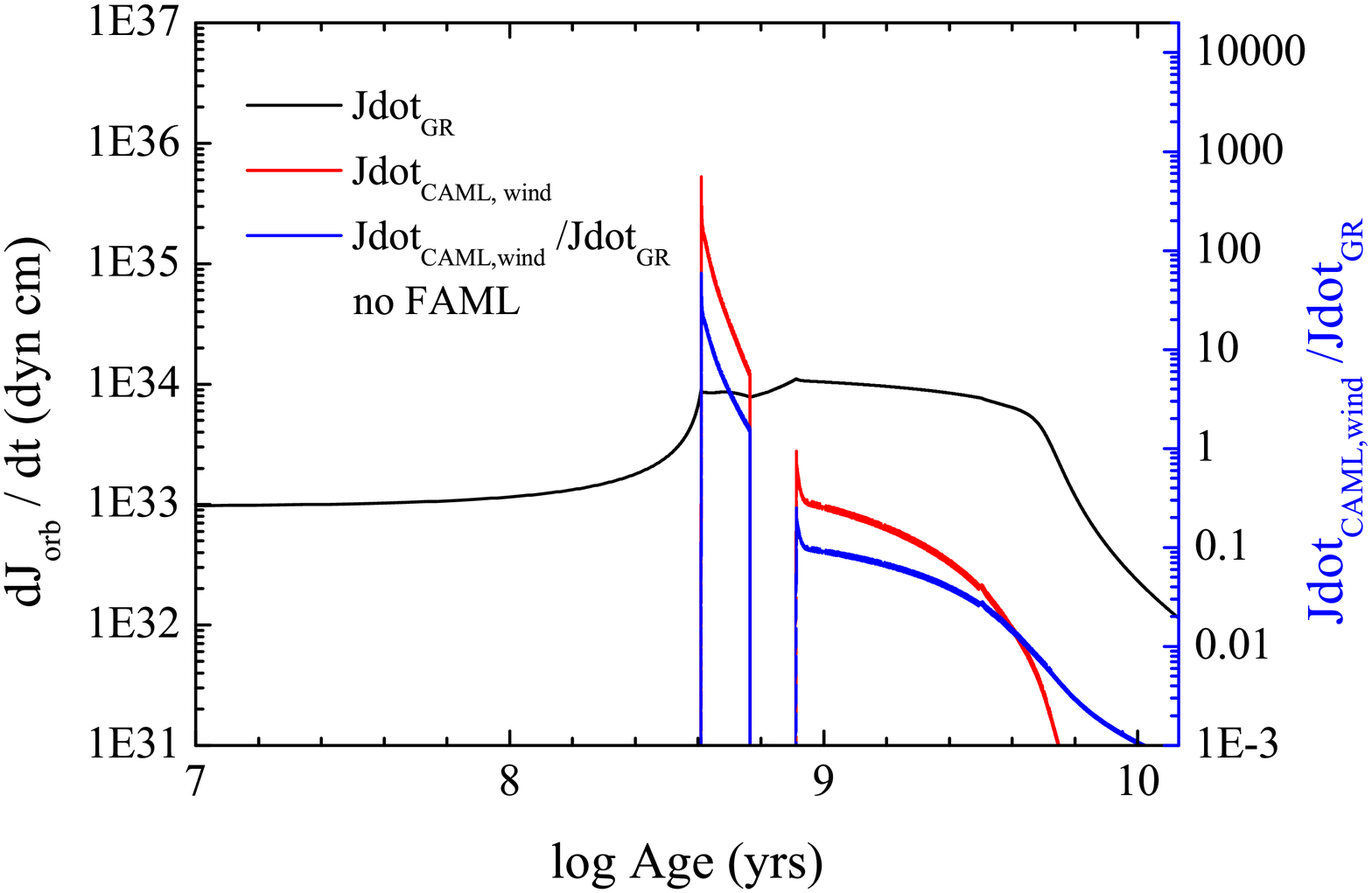}
\includegraphics[scale=0.25]{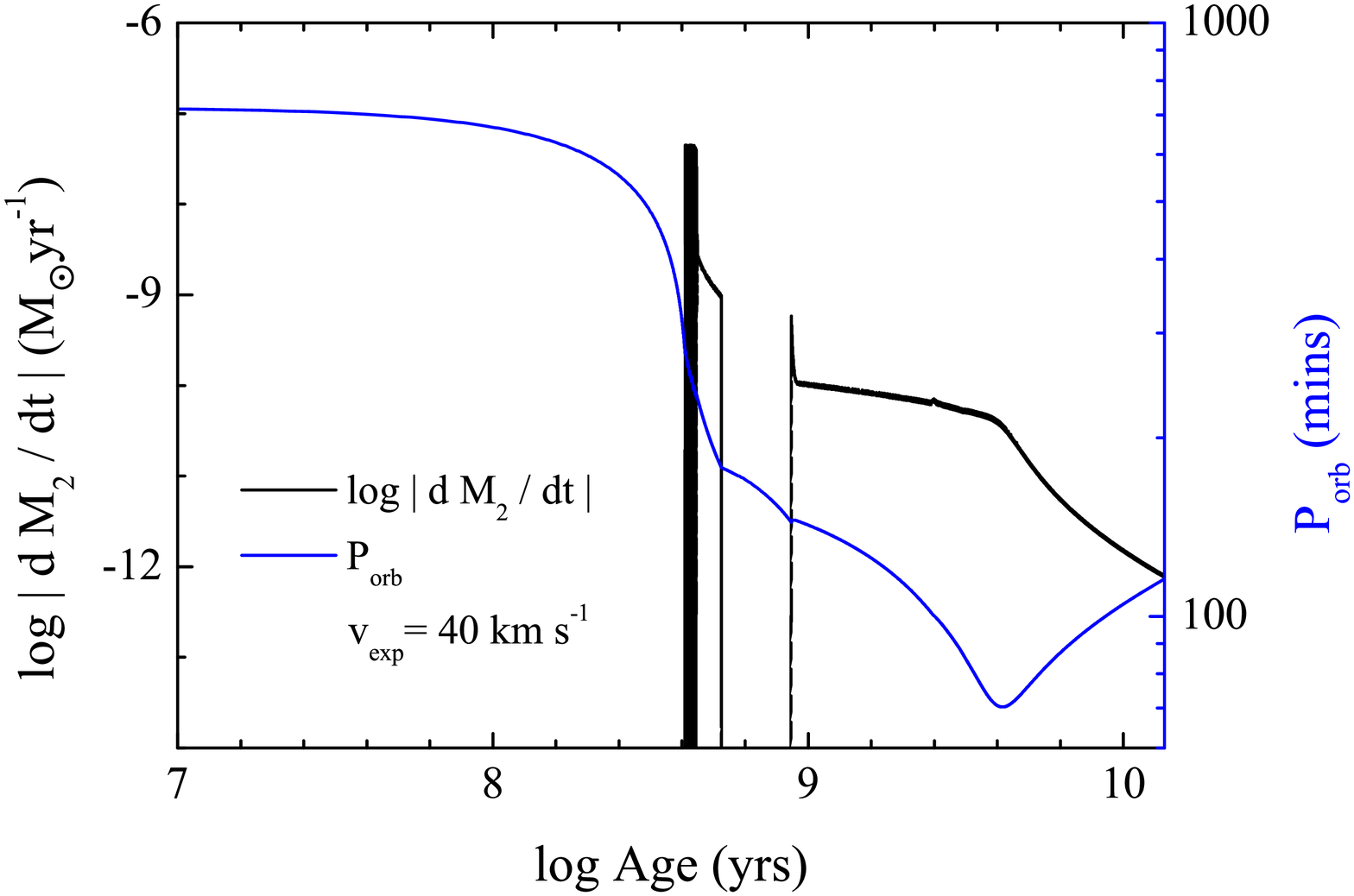}\includegraphics[scale=0.25]{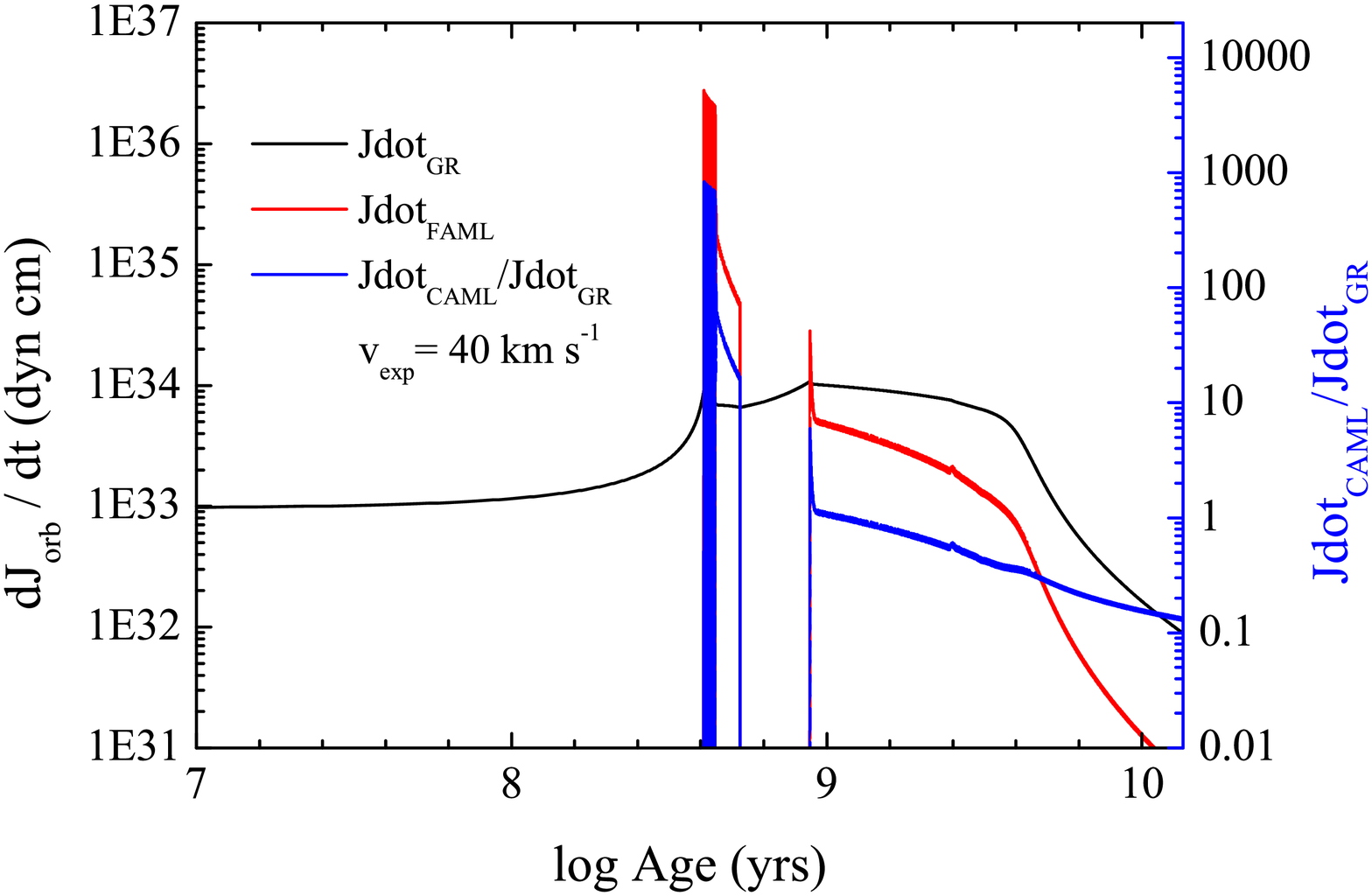}
\includegraphics[scale=0.25]{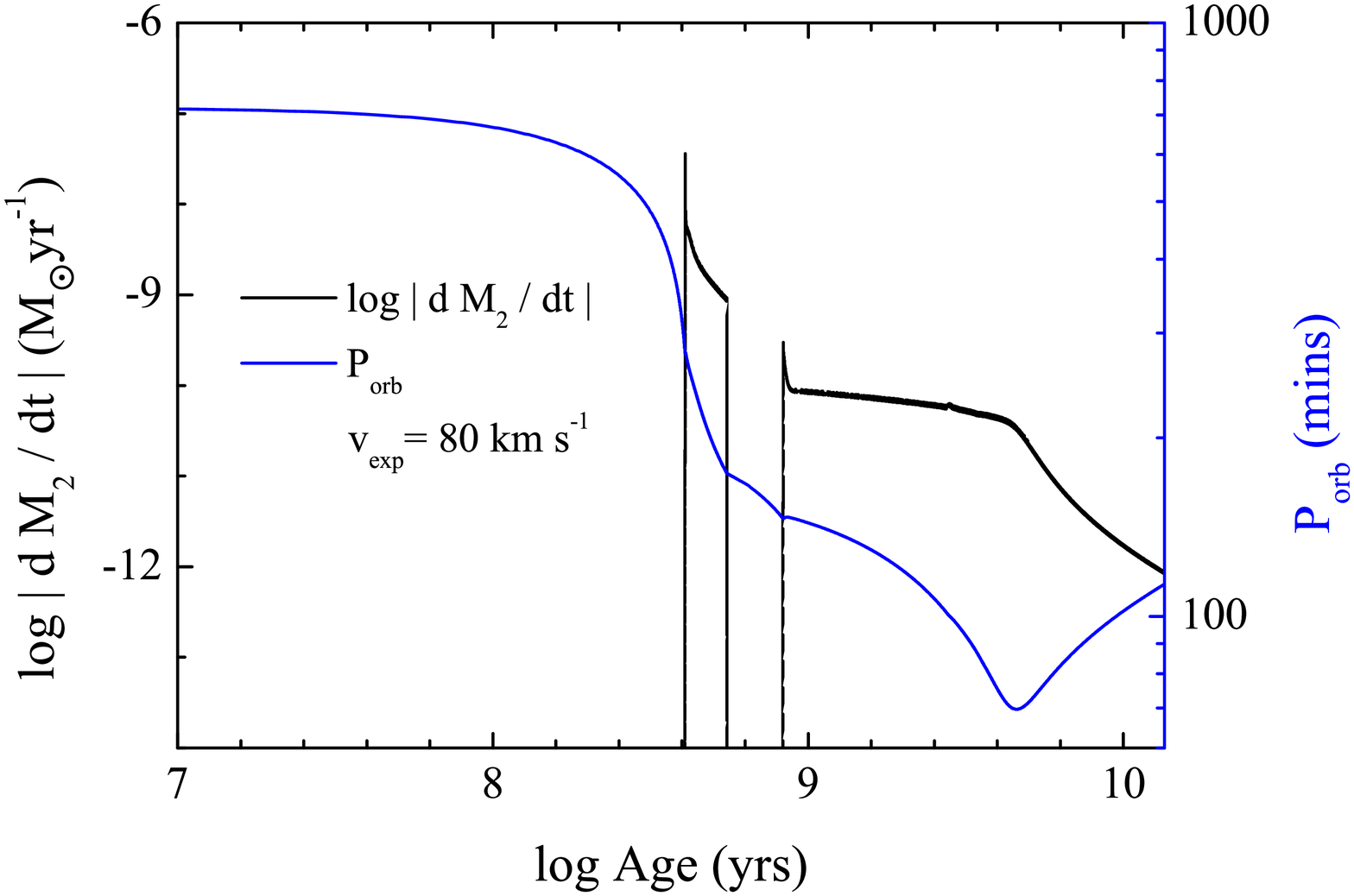}\includegraphics[scale=0.25]{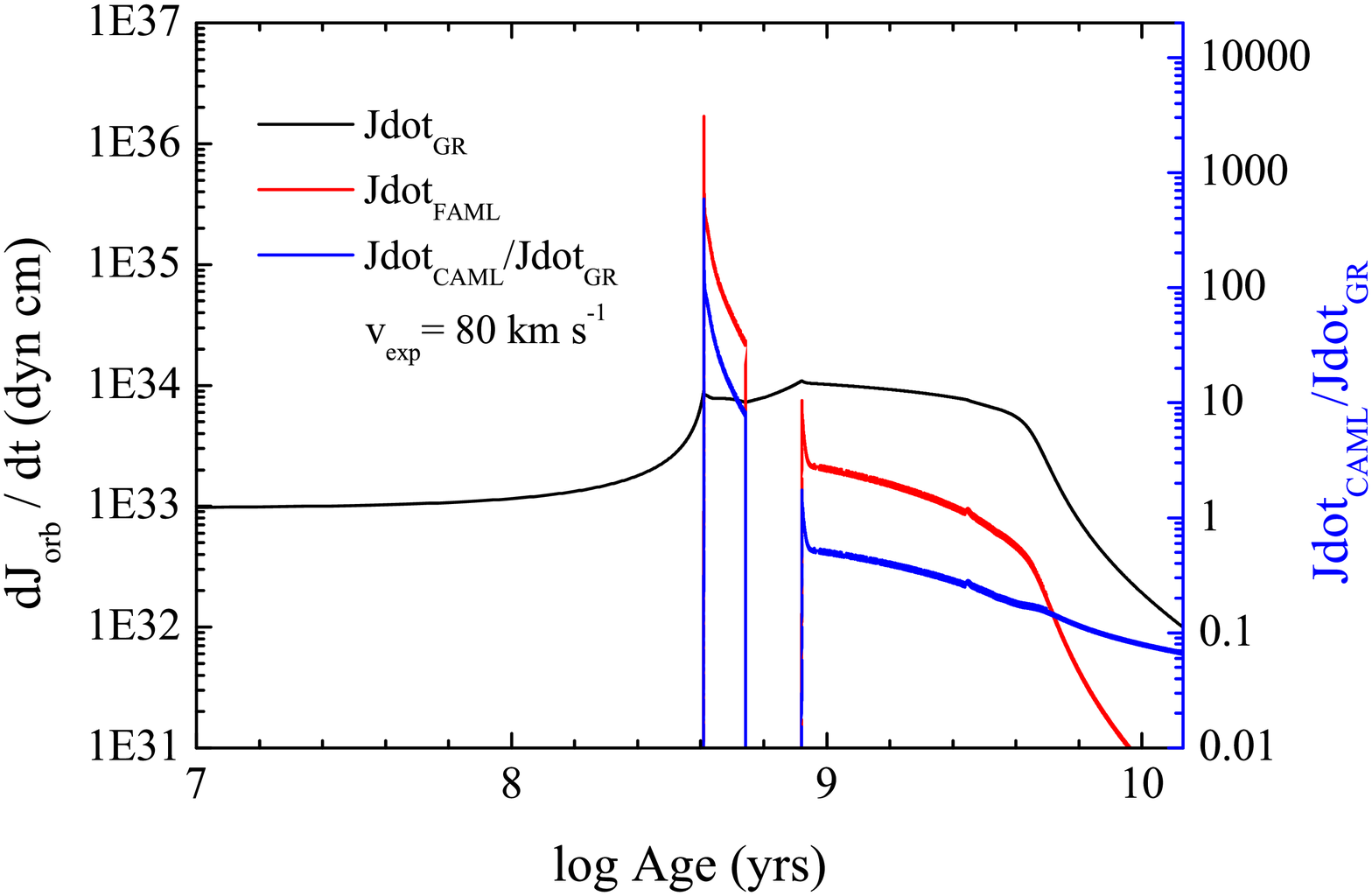}
\includegraphics[scale=0.25]{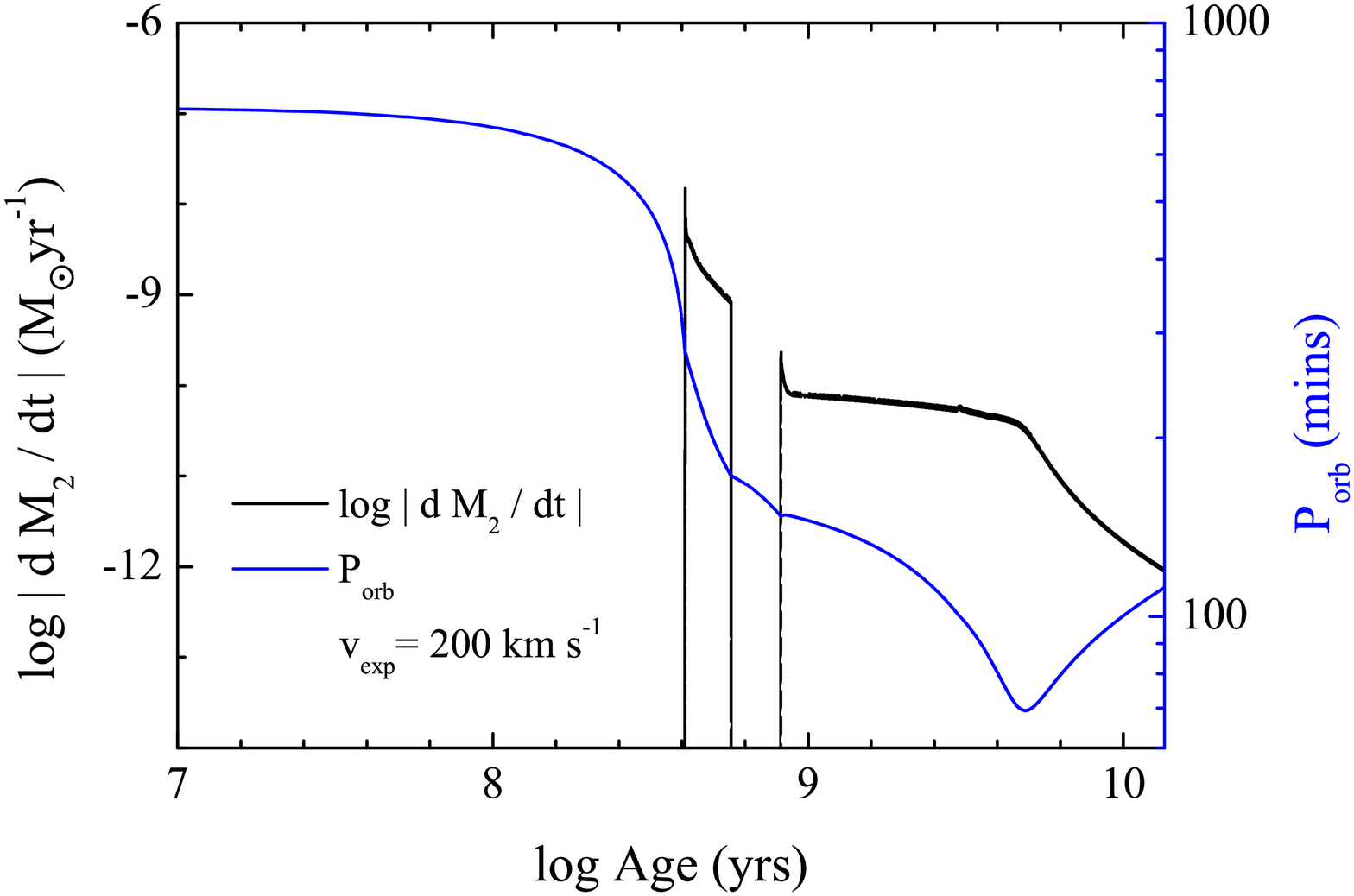}\includegraphics[scale=0.25]{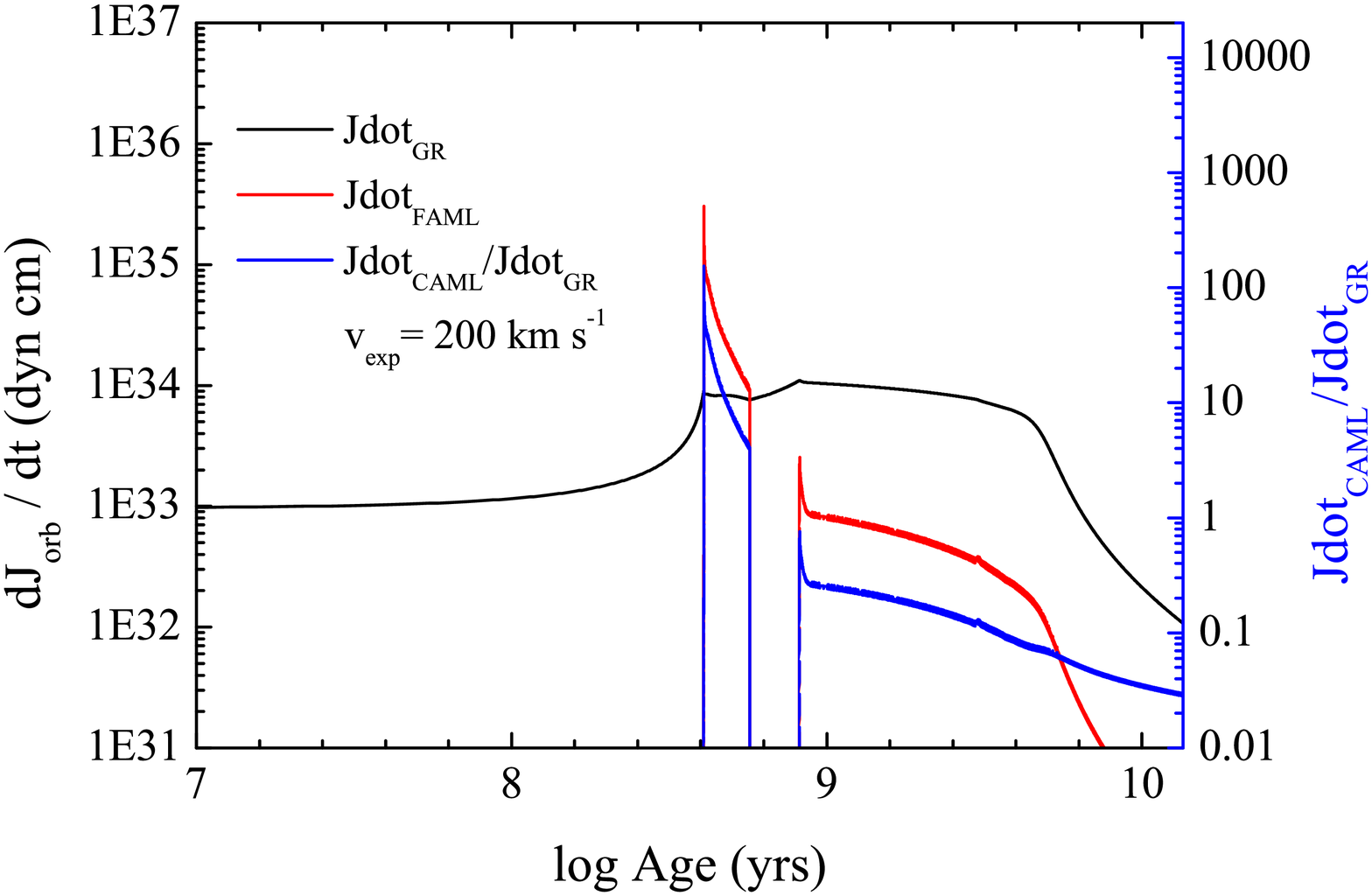}
\caption{Same as Fig.~1, but with $M_{\rm {2,i}}$ = 0.6 $M_\odot$, $M_{\rm {WD,i}}$ = 0.8 $M_\odot$ and $P_{\rm orb,i}$ = 0.5 day.}
\label{fig:subfig}

\end{figure}

\clearpage

\begin{figure}
\centering
\includegraphics[scale=0.25]{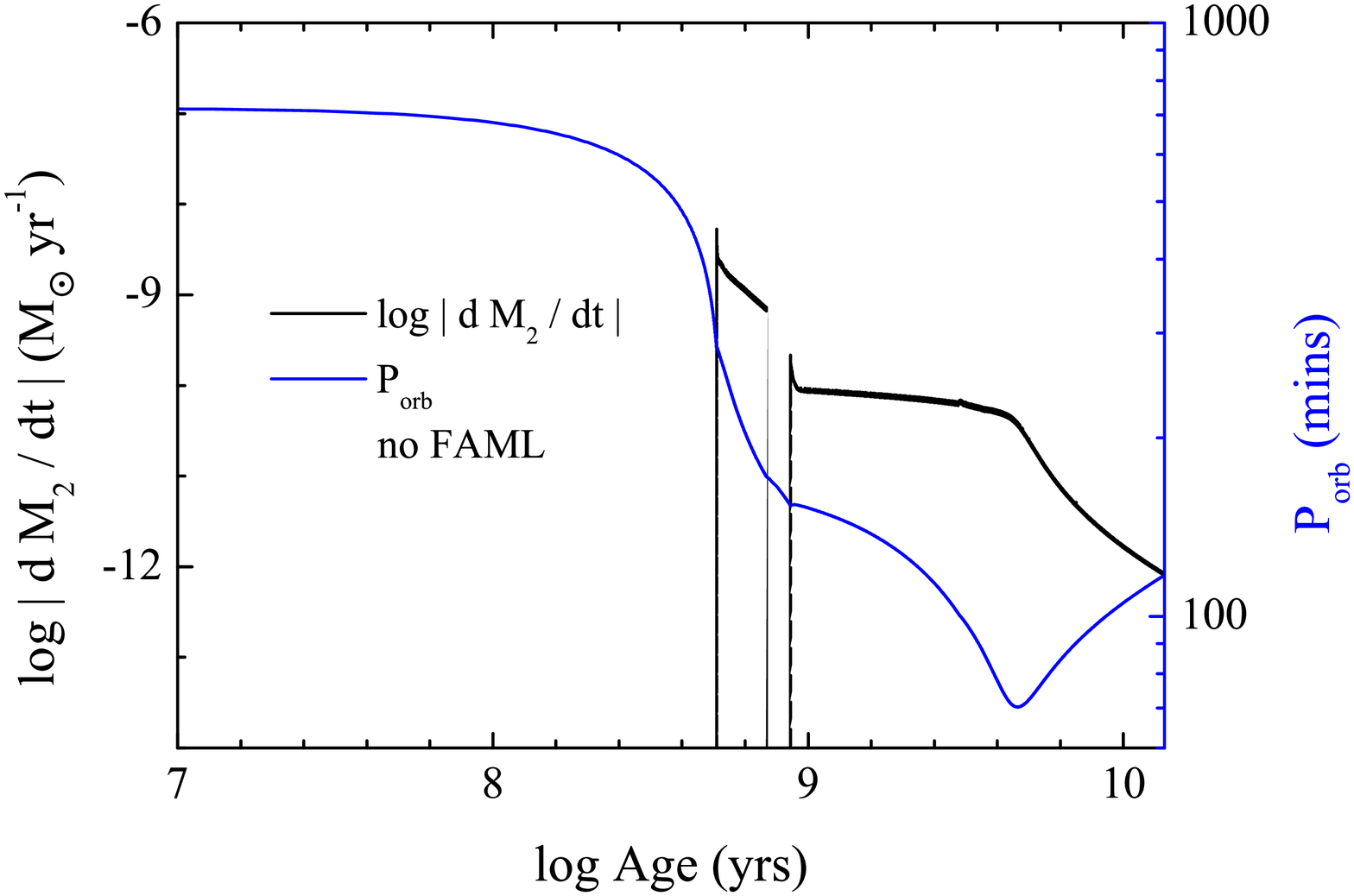}\includegraphics[scale=0.25]{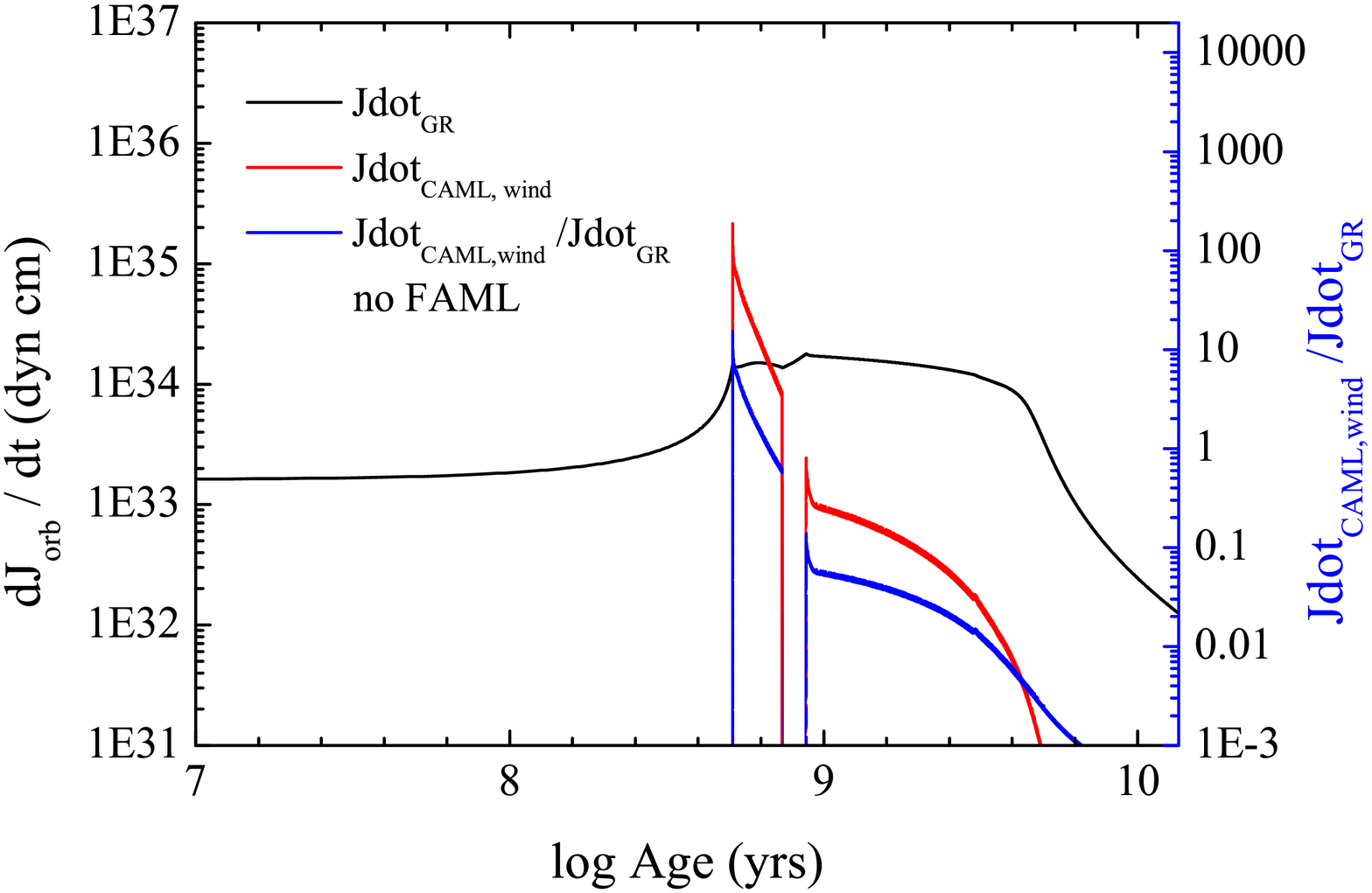}
\includegraphics[scale=0.25]{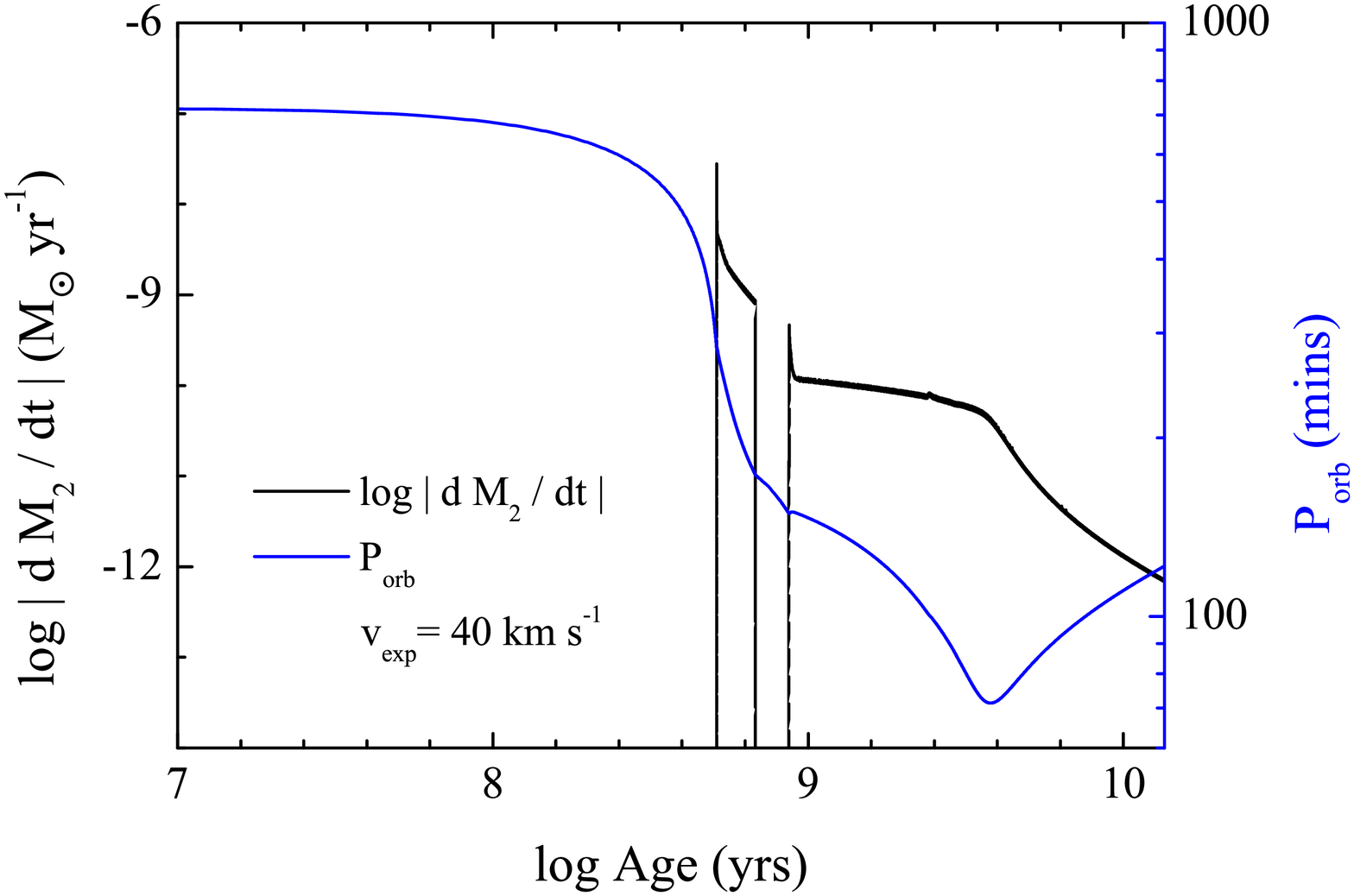}\includegraphics[scale=0.25]{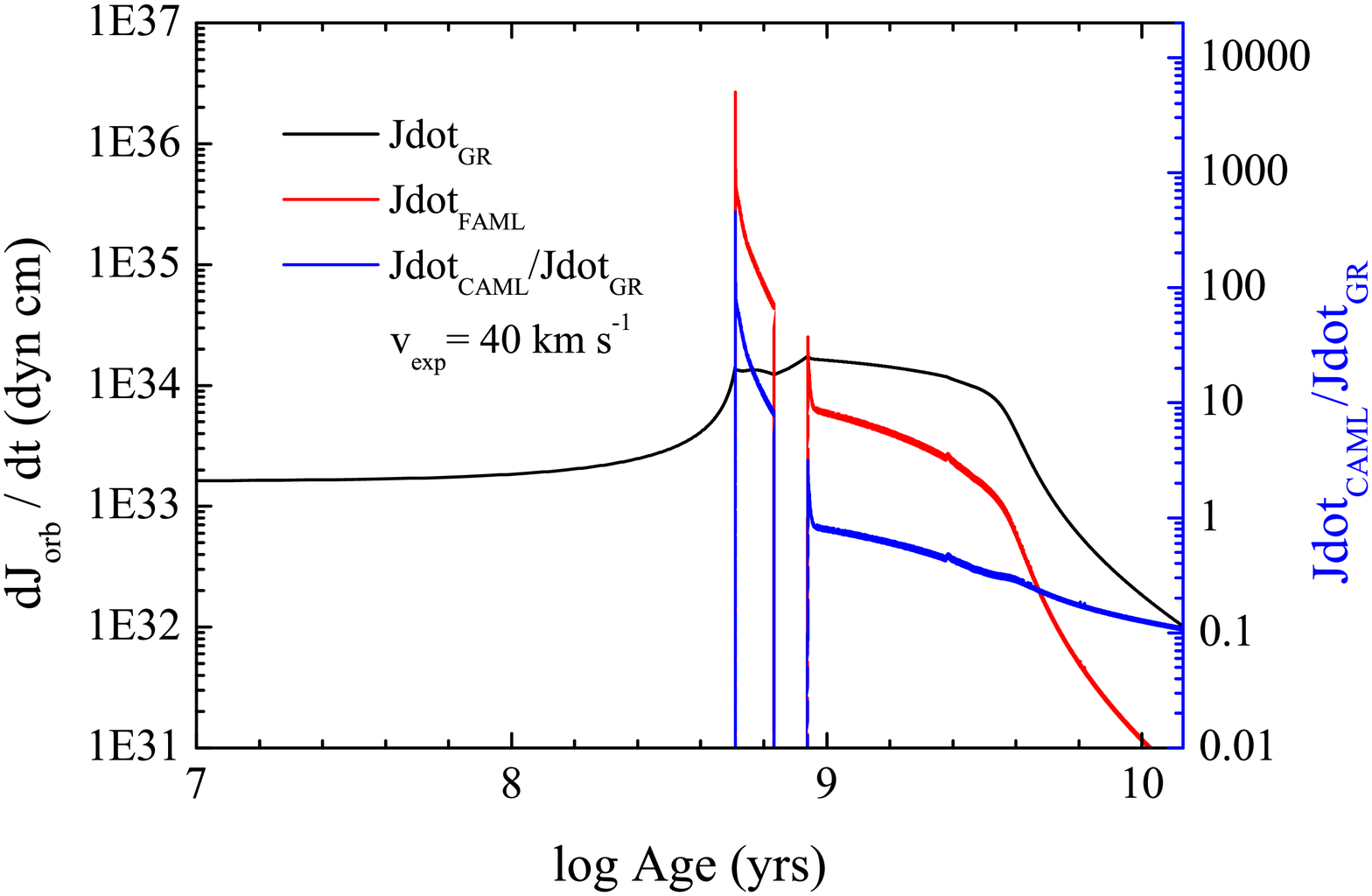}
\includegraphics[scale=0.25]{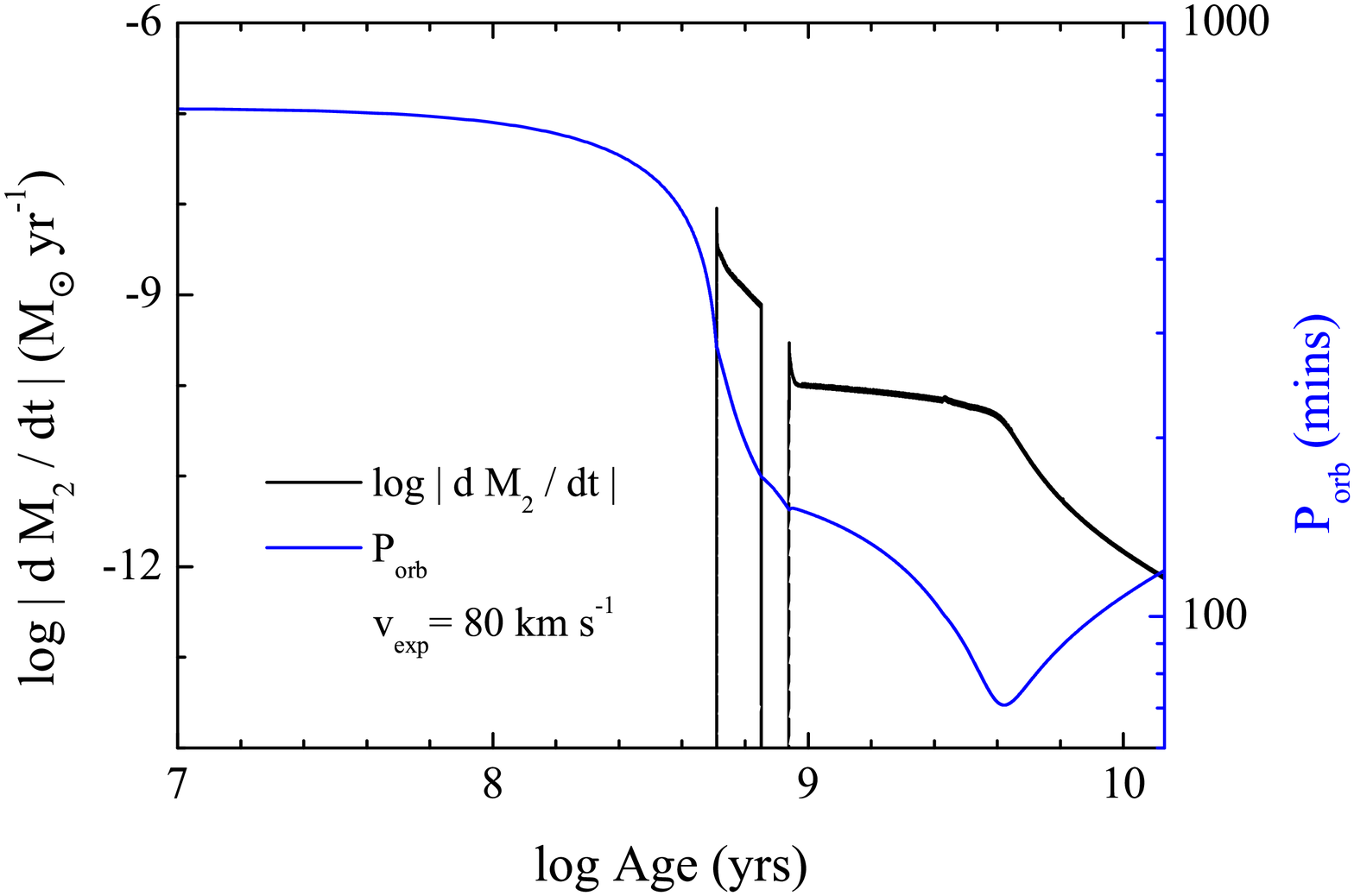}\includegraphics[scale=0.25]{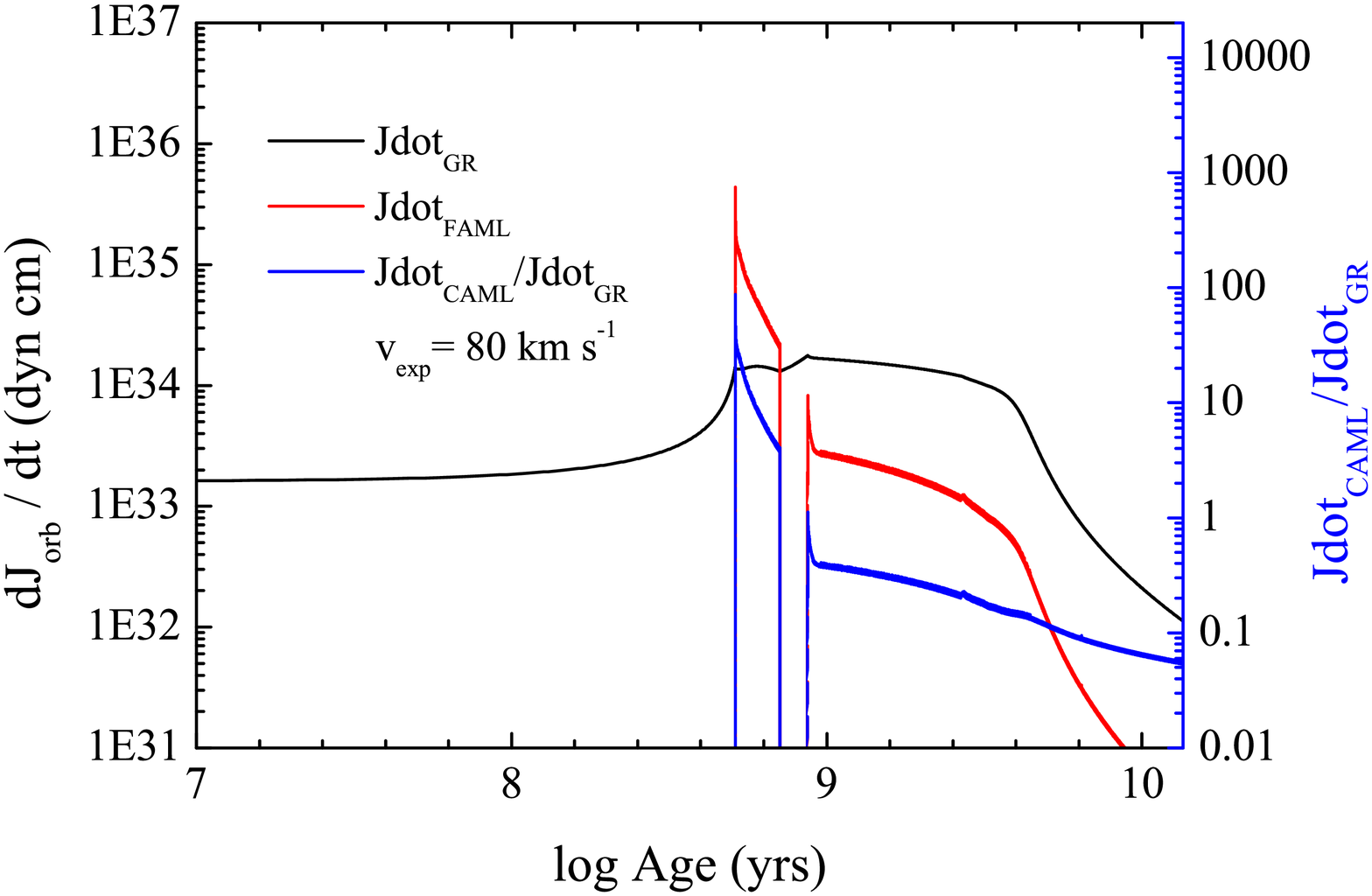}
\includegraphics[scale=0.25]{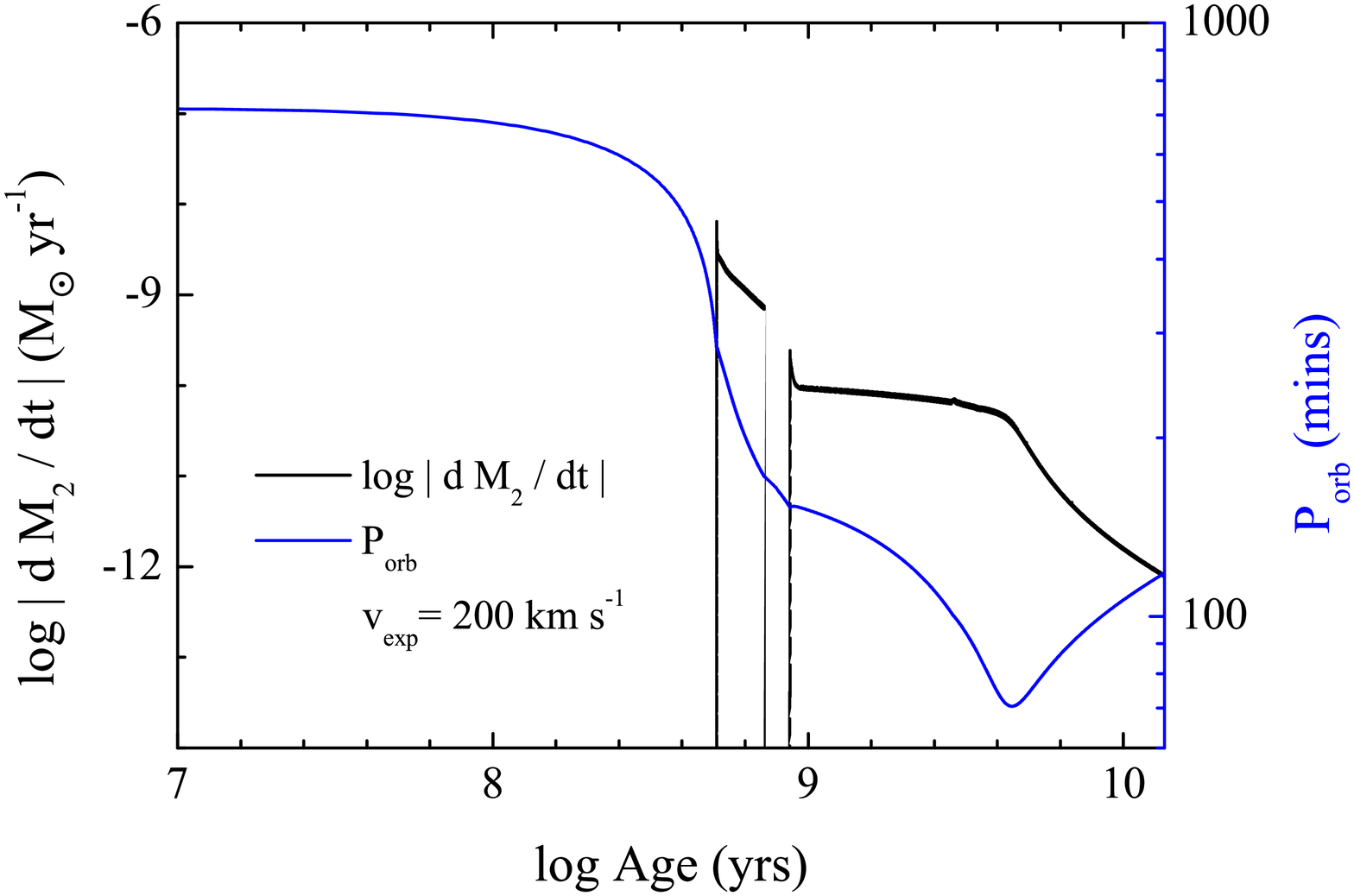}\includegraphics[scale=0.25]{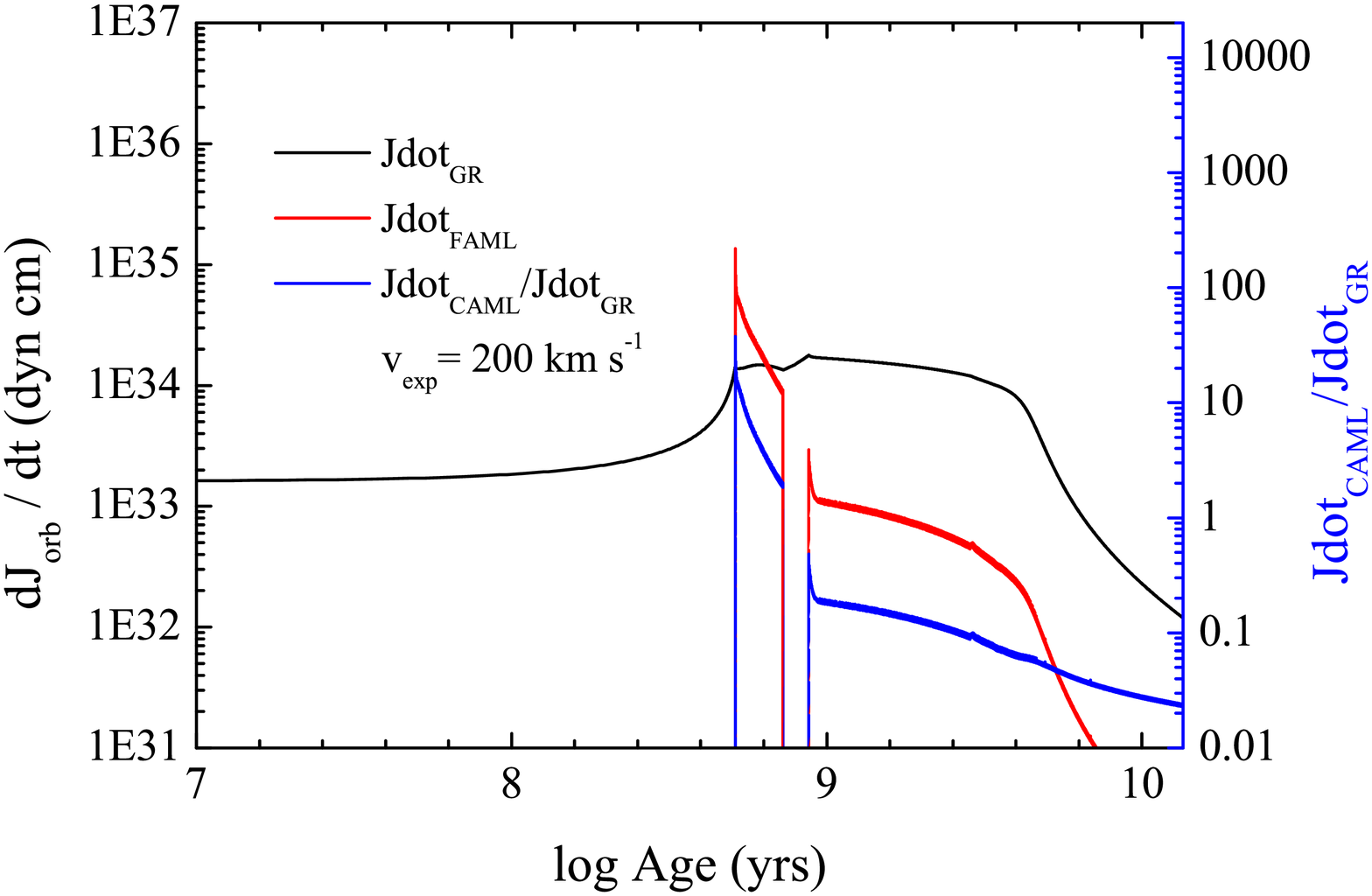}
\caption{Same as Fig.~1, but with $M_{\rm {2,i}}$ = 0.6 $M_\odot$, $M_{\rm {WD,i}}$ = 1.1 $M_\odot$ and $P_{\rm orb,i}$ = 0.5 day.}
\label{fig:subfig}

\end{figure}

\clearpage

\begin{figure}
\centering
\includegraphics[scale=0.35]{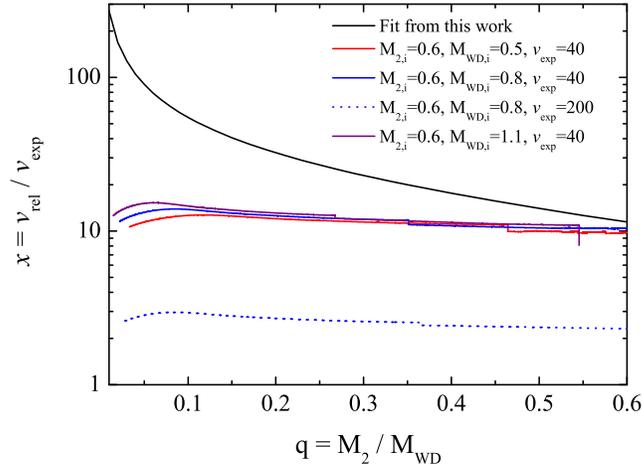}
\caption{The black line shows the predicted relation between $q$ and $x$ if FAML can account for the extra AML below the period gap in CVs. Other lines represent the calculated $x-q$ relations in CV evolution. The units of the mass and the expansion velocity are $M_\odot$ and $\rm {km\,s}^{-1}$, respectively. }
\label{fig:subfig}

\end{figure}

\clearpage

\label{lastpage}

\end{document}